\documentclass[a4paper,superscriptaddress,twocolumn]{revtex4}
\usepackage{amssymb}
\usepackage{amsfonts}
\usepackage{amsmath}
\usepackage{graphicx}
\usepackage{dcolumn}
\usepackage{bm}
\usepackage[english]{babel}
\usepackage[utf8]{inputenc}
\usepackage{comment}
\usepackage[sort&compress]{natbib}
\usepackage{hyperref}
\hypersetup{colorlinks=true,citecolor=blue,linkcolor=blue,urlcolor=black}
\usepackage[colorinlistoftodos]{todonotes}
\usepackage{cancel}
\usepackage{fourier-orns}

\begin{document}
\title{Exact results on analogue gravity in optical Plebanski-Tamm media}
%\maketitle

\author{Franco Fiorini}
\email{francof@cab.cnea.gov.ar} 

\affiliation{Consejo Nacional de Investigaciones Cient\'ificas y T\'ecnicas (CONICET), S. C. de Bariloche, Río Negro 8400, Argentina.}
\affiliation{Depto. de Ingenier\'ia en Telecomunicaciones, Centro At\'omico Bariloche, Comisi\'on Nacional de Energ\'ia At\'omica, Av. Ezequiel Bustillo 9500, S. C. de Bariloche, R\'io Negro 8400, Argentina.}
\affiliation{Instituto Balseiro, Universidad Nacional de Cuyo, Av. Ezequiel Bustillo 9500, S. C. de Bariloche, Río Negro 8400, Argentina.}

\author{Santiago M. Hernandez}
\email{shernandez@ib.edu.ar} \affiliation{Instituto Balseiro, Universidad Nacional de Cuyo, Av. Ezequiel Bustillo 9500, S. C. de Bariloche, Río Negro 8400, Argentina.}
\author{Edith L. Losada}
\email{losada@cab.cnea.gov.ar} 
\affiliation{Depto. de Ingenier\'ia en Telecomunicaciones, Centro At\'omico Bariloche, Comisi\'on Nacional de Energ\'ia At\'omica, Av. Ezequiel Bustillo 9500, S. C. de Bariloche, R\'io Negro 8400, Argentina.}

\begin{abstract}

Exact results concerning ray-tracing methods in Plebanski-Tamm media are derived. In particular, Hamilton equations describing the propagation of quasi-plane wave electromagnetic fields in the geometrical optics regime are explicitly written down in terms of the 3-metric representing the properties of the optical analogue, anisotropic medium. We exemplify our results by obtaining the trajectories of light in the resulting analogue medium recreating G\"odel's universe.
\end{abstract}
\maketitle
\section{Introduction}

General Relativity (GR) is transiting a reinvigorating stage nowadays, quite comparable to the one experienced in its best moments. This is particularly evident from an observational point of view, where experiments, once only imagined, are currently providing detailed information about the behavior of the gravitational field in strong curvature regions very distant in time and space, where extreme gravitational events are unfolding. Actually, it is practically unanimous the fact that black holes are much more common objects than initially expected, not only in far galaxies \cite{Event}, but also in our own \cite{Ghez}. Sharing this compelling observational evidence, gravitational wave astronomy is becoming a quite prolific experimental branch as well, specially since the measurement of gravitational waves emitted by black hole mergers was achieved \cite{Abbot}.

However, this optimistic experimental scenario has intrinsic limitations. For instance, quantum field effects in curved spacetime backgrounds, as Hawking radiation, are so extremely small for black holes of astrophysical size, that any attempt to measure them is practically inconceivable. On the other hand, the extreme energies characterizing the gravitational field in the vicinity of black hole singularities, are basically causally disconnected from us, shielded by event horizons. Curiously, these limitations have launched a new area of research in which the main concern is the possibility of reproducing or emulating in the lab, to a certain extent, the conditions to favor the existence of such elusive effects. This is the aim of the so called analogue gravity models \cite{Analog1}, \cite{Tanos}. Among the newest conceptual frameworks in the area, optical analogue models were developed mainly in the last decade, see e.g., \cite{leon-phil}, \cite{leon-phi2}. They are based, in turn, on quite established principles coming from the study of electromagnetic fields on curved spacetimes. For instance, and quite surprisingly, the optical analogue of the Hawking radiation was reported very recently within the context of nonlinear optics \cite{RAD}.

Perhaps even more striking than the existence of black hole solutions in GR, is the possibility of time travel or, equivalently, faster-than-light space travel. This is actually not that uncommon among GR solutions. Some of them require the existence of negative energy density, as in the case of some traversable Lorentzian wormholes \cite{Morris}, but causal violations are also present in solutions with perfectly realistic energy-momentum tensors, as in the Reissner-Nordstr\"{o}m electrically charged black hole and Godel's cosmological solution, being the latter central in this work. Time travel is also possible as a consequence of having GR solutions in pure vacuum, as in the Kerr rotating black hole spacetime and in Gott's moving cosmic strings solution \cite{Gott}. However, it is fair to say that the formation of Cauchy horizons and the creation, then, of regions with causal violations, do not seem to be favored in Nature because of the so-called Chronology Protection Conjecture, at least in the case of compactly generated Cauchy horizons \cite{Rengo1}. Nonetheless, the Kerr and Reissner-Nordstr\"{o}m spacetimes are good examples of solutions with non-compactly generated Cauchy horizons. 

The purpose of this work is to further characterize the propagation of quasi-plane waves in anisotropic optical media by using the non-covariant Plebanski-Tamm (PT) constitutive relations, and to apply the results to the optical analogue of G\"{o}del's universe. According to this  formalism, the propagation of light in a given curved spacetime (not necessarily a solution of Einstein's equations), can be viewed as an optical problem in a flat anisotropic material medium with rather peculiar electromagnetic properties. This analogy enables to link the study of null geodesics of a given (3+1)-dimensional curved manifold with the characterization of light paths in a 3-dimensional Euclidean space filled with a PT medium. In this way one can, in principle, mimic in the lab curved spacetime effects, provided one has the ability to construct such a peculiar material. Fortunately, the relatively recent appearance of metamaterials is helping to shorten the gap between the physics in the lab and the one occurring in distant regions of the Universe; the development of new and interesting metamaterials exhibiting rare optical properties is an ongoing and flourishing activity, see, e.g.,  Refs.~\cite{Meta1}-\cite{Meta3}.

It is well known that G\"{o}del's solution does not provide a realistic description of the Universe; however, it offers a relatively simple example of existence of closed causal curves, i.e., of causal violations. It results important, then, to inquire on how this causal pathologies influence the behavior of the analogue optical system. Needless to say, the presence of closed null or timelike curves is a purely (3+1)-spacetime phenomenon, utterly absent in any 3-dimensional (merely spatial) description, as the one concerning PT media. Quite often, however, the strongly warped spacetime structure responsible for the presence of closed causal curves is also the culprit of a very intricate behavior in the ``usual'' geodesic curves. The corresponding intricate behavior of light in the optical analogue is what primarily motivates this work. 

This article is organized as follows: in section \ref{sec2} we introduce the necessary established material regarding G\"{o}del's solution, and the rudiments of the electrodynamics in PT media. Even though we have kept this section as short as possible, we felt that some details could be helpful; these are contained in Appendix \ref{ap1}. Our contribution to the subject is contained in sections \ref{sec3} and \ref{sec3c}. In \ref{sec3a}, a discussion is carried out in regard to the constrained Hamiltonian system for ray tracing in PT media. There we stress a very important point which seems to have gone unnoticed in the literature, i.e., the fact that the dispersion relation is not naturally incorporated in the evolution equations. In this section, Hamilton's equations for light rays are explicitly derived in terms of the optical properties of the PT medium. In \ref{sec3b} the optical analogue of G\"odel's solution is worked out, and the Hamilton's equations for light rays presented in \ref{sec3a} are exactly solved. Finally, in \ref{sec3c} the solutions are fully analyzed and discussed.

Because this work involves $(3+1)$-dimensional spacetime objects as well as standard 3-dimensional ones, a word of caution regarding the notation will be helpful. Vector and second rank tensor components will be denoted using Greek indices $v_{\mu}$ and $g_{\mu\nu}$, respectively, where $\mu,\nu:0,1,2,3$, or Latin indices $v_{i}$ and $g_{ij}$ if $i,j:1,2,3$. Three dimensional vector objects, will in turn be denoted by $\bar{v}$, and their scalar, vector and tensor products as $\bar{v}\cdot\bar{w}$, $\bar{v}\times \bar{w}$ and $\bar{v}\otimes\bar{w}$, respectively. However, for operational reasons, scalar products of row and column vectors will be written without dots, as in $\bar{p}^{\intercal}\bar{q}$, where $^{\intercal}$ means transposition.  Moreover, 3D-matrices will be written as $\textbf{M}$. Cartesian vector and matrix components will be written $a_{i}$ and $M_{ij}$, respectively. The product of a matrix $\textbf{M}$ by a vector $\bar{v}$ will be simply $\textbf{M}\bar{v}$. Einstein's summation convention will be used when needed also in three dimensions. All the components in 3D will be referred to a Cartesian coordinate system $\bar{x}=(x_{1},x_{2},x_{3})$.

\begin{center}
   \large $\ast \; \ast \; \ast$   
\end{center}

\vfill

\section{Preliminary material}\label{sec2}

\subsection{G\"odel's spacetime}\label{sec2a}

In units where $c=(\epsilon_{0} \mu_{0})^{-1/2}=1$, the line element of the G\"odel solution \cite{Godel} can be described in local $(t,\bar{x})$ coordinates as \cite{hawking_ellis}:
\begin{equation}\label{metgSI}
	ds^2= - \big(dt + e^{\sqrt{2}\,\omega x_{1}} dx_{2}\big)^2 + dx_{1}^2 +\frac{e^{2\sqrt{2}\,\omega x_{1}}}{2}  dx_{2}^2+ dx_{3}^2,
\end{equation}
where $\omega$ is a constant parameter with units of inverse length. Metric (\ref{metgSI}) is a solution of Einstein's field equations
\begin{equation}\label{Einstein}
	R_{\mu\nu}-\frac{1}{2}\,R\, g_{\mu\nu}+\Lambda g_{\mu\nu} =8 \pi T_{\mu\nu},
\end{equation}
for a pressure-free perfect fluid, i.e., if $T_{\mu\nu}$ is of the form  $T_{\mu\nu}=\rho\, v_\mu v_\nu$, where $\rho$ is the energy density and $v_\mu=\delta^{0}_{\mu}$ is the four-velocity vector of the flow. The parameters of the solution are related according to

\begin{equation}\label{met}
4\pi \rho =\omega^2=-\Lambda,
\end{equation}
hence G\"odel's solution involves a constant positive energy density and a negative cosmological constant. As a consequence of this, G\"odel's universe turns out to be a spacetime of constant positive scalar curvature $R$ proportional to $\omega^{2}$.

G\"odel's solution has a number of peculiarities which makes it interesting from a fundamental point of view. On the one hand, the spacetime is geodesically complete; geodesics can be extended to arbitrary values of the affine parameter, and as a consequence of this, the spacetime is singularity free. Geodesically complete spacetimes are rare among Einstein's GR solutions, and they are more the exception than the rule. On the other hand, the solution contains closed causal (non geodesic) curves, announcing causal pathologies of different sorts. These are not evident from the form (\ref{metgSI}) of the metric; however, a change of coordinates will bring them to light. Actually, defining new (non-dimensional) coordinates $(t',r,\phi,x_3')$ according to 

\begin{align}
	&\exp (\sqrt{2}\,\omega\, x_1)=\cosh(2r)+\cos\phi\sinh(2r)\label{change1}\\[0.3cm]
	&\omega\,x_2\exp (\sqrt{2}\,\omega x_1)=\sin\phi\sinh(2r)\label{change2}\\[0.3cm]
	&\sqrt{2}\,\omega\, x_3 = x_3'\\[0.3cm]
	&\tan\Big[\resizebox{0.3cm}{!}{$\frac{1}{2}$}\big(\phi + \omega t -\sqrt{2}\,t'\big)\Big]=\exp (-2r)\tan(\phi/2),\hspace{0.3cm}
	\end{align}
	the metric (\ref{metgSI}) is transformed into the following conformal expression 
	\begin{eqnarray}
	ds^2&=&\frac{2}{\omega ^2} \Big[-dt'^{\,2} + dr^2 + \frac{1 - \sinh ^2(r)}{2} \sinh ^2(r)\, d\phi^2 \nonumber \\
	&&-2 \sqrt{2}\, \sinh ^2(r) dt' d\phi + dx_3'^2 \Big] \,.\label{metg2}
	\end{eqnarray}
If we fix constant values of $t'$, $r$ and $x_3'$, the interval (\ref{metg2}) reduces to
\begin{equation}\label{ctcs}
ds^2=\omega^{-2}\sinh^2(r)(1 - \sinh^2(r))d\phi^2.
\end{equation}
After the identification of $\phi=0$ with $\phi=2\pi$, the null curve leading to $ds^2=0$ is a closed null curve. This curve verifies

\begin{equation}\label{ctcs2}
\sinh^2(r)-1=0,\,\,\,\,\,\,r=r_{0}= \ln(1+\sqrt{2}).
\end{equation}
Furthermore, curves with constant $t'$, $r$ and $x_3'$ values are closed timelike curves provided $r>r_{0}$, because the interval (\ref{ctcs}) becomes negative. Hence, the closed null curve defined by $r=r_0$ constitutes the boundary of the region where causal violations are admissible. However none of these closed curves are actually geodesics of the G\"odel spacetime, but merely closed curves. 

In turn, null geodesics spiral endlessly; for instance, starting from a given event $q$ at $r=0$, they diverge to reach a maximum radius $r=r_{0}$, and then reconverge to $r=0$ in the future of $q$. This behavior will find a correlate in the analogue optical medium to be dealt with in section \ref{sec3b}. Details on the structure of the G\"odel spacetime can be found in \cite{Tedescos} \cite{Tedescos2}.

\subsection{Geometrical optics in Plebanski-Tamm media}\label{sec2b}

Transformation optics is about the properties of light in rather unusual (meta) material media. The basis of the analogy between the propagation of light in an arbitrary (3+1)D-curved spacetime described in local coordinates by the metric tensor $g_{\mu\nu}$, and its corresponding propagation in a flat 3D-space filled with a very peculiar material, is given, within the context of the most simple approach, by Plebanski's constitutive equations. For details, we refer the reader to the original work \cite{Pleb} and to the abundant, more recent developments in the field (see, e.g. \cite{leon-phil} and \cite{leon-phi2}). The aforementioned analogy relies on the constitutive relations
\begin{align}
\bar{D}(\bar{x},t) =  \textbf{K}(\bar{x})\,\bar{E}(\bar{x},t) - \bar{\Gamma}(\bar{x})\times\bar{H}(\bar{x},t)\label{eq:ecconstitD},\\%[.3cm]
\bar{B}(\bar{x},t) =  \textbf{K}(\bar{x})\,\bar{H}(\bar{x},t) + \bar{\Gamma}(\bar{x})\times\bar{E}(\bar{x},t), \label{eq:ecconstitB}
\end{align}
where the components of the matrix $\textbf{K}$ and the vector $\bar{\Gamma}$ are related to the spacetime metric according to
\begin{align}
K_{ij}&=-\frac{\sqrt{-det(g_{\mu\nu})}}{g_{00}}g^{ij}\label{relgama}\\
\Gamma_{m}&=\frac{g_{0m}}{g_{00}}.\label{elotrogama}
\end{align}
Equations~(\ref{eq:ecconstitD})-(\ref{eq:ecconstitB}) seem to have been obtained previously by Tamm \cite{Tamm}, so it seems fair to refer to such media as Plebanski-Tamm (PT) media.

The fine expressions (\ref{eq:ecconstitD})-(\ref{eq:ecconstitB}) involve two drawbacks; on the one hand, the matrix $\textbf{K}$ officiates in (\ref{eq:ecconstitD}) as a relative permitivity tensor, and in (\ref{eq:ecconstitB}) as a relative permeability. PT media are then characterized by equal relative permitivity and permeability. Even though this could seem very strange, the advent of metamaterials is making possible the design and manufacture of materials with very bizarre electromagnetic properties. The equality, to a certain extent, of the electric and magnetic properties of a material is certainly not that far from being achievable.

On the other hand, PT equations (\ref{eq:ecconstitD})-(\ref{eq:ecconstitB}) are explicitly non covariant; they are valid only in a cartesian coordinate system fixed in the laboratory frame held at rest in the stationary medium. Moreover, the expression (\ref{relgama}) is dubious from a mathematical point of view, because it relates a second rank contravariant tensor field with the object $K_{ij}$ which is not even a covariant tensor. However, covariant extensions to general (non stationary) media were developed (see, e.g.  \cite{DeFelice}, \cite{PlebCOV}), and it was shown that PT equations are completely equivalent to the more general covariant approach for the case of a stationary medium. PT equations are then perfectly suitable in order to emulate the effects of a curved spacetime in the lab.  

\bigskip

Geometrical optics in PT media involves quasi-plane wave electromagnetic fields of the form

\begin{eqnarray}
\bar{E}(\bar{x},t)&=&\bar{E}_{0}(\bar{x})\exp\big[i\,k_{0}\,(\bar{k}(\bar{x})\cdot\bar{x}- t)\big],\label{expEyH1}\\
\bar{H}(\bar{x},t)&=&\bar{H}_{0}(\bar{x})\exp\big[i\,k_{0}\,( \bar{k}(\bar{x})\cdot\bar{x}- t)\big].\label{expEyH2}
\end{eqnarray}
Here, $\bar{E}_{0}(\bar{x})$ and $\bar{H}_{0}(\bar{x})$ are the space-dependent complex-valued field amplitudes and the vector $\bar{k}(\bar{x})$ is the non dimensional relative wave vector. Note that in (\ref{expEyH1}) and (\ref{expEyH2}), both $\bar{x}$ and $t$ have units of length, and $k_{0}$ of inverse length; this is a consequence of the fact that $c=1$. Hereafter, we shall omit the explicit space and time dependence in the expressions in question. 

Combining the constitutive relations (\ref{eq:ecconstitD})-(\ref{eq:ecconstitB}) and the ansatz (\ref{expEyH1})-(\ref{expEyH2}) and evaluating the fields in the source-free Maxwell curl equations

\begin{eqnarray}\label{mostro}
\bar{\nabla}\times \bar{E}=-\frac{\partial \bar{B}}{\partial t},\,\,\,\,\bar{\nabla}\times \bar{H}=\frac{\partial \bar{D}}{\partial t},
\end{eqnarray}
we obtain
\begin{eqnarray}
i\,k_{0}^{-1}\,\bar{\nabla}\times\bar{E}_{0}&=&(\bar{\nabla}(\bar{k}\cdot\bar{x})-\bar{\Gamma})\times\bar{E}_{0}-\textbf{K} \,\bar{H}_{0}\notag\\
i\,k_{0}^{-1}\,\bar{\nabla}\times\bar{H}_{0}&=&(\bar{\nabla}(\bar{k}\cdot\bar{x})-\bar{\Gamma})\times\bar{H}_{0}+\textbf{K} \,\bar{E}_{0}\notag
\end{eqnarray}
Besides the quasi-plane wave character of the fields (\ref{expEyH1})-(\ref{expEyH2}), geometrical optics requires slow variations of the constitutive fields $\bar{E}$, $\bar{H}$ and $\bar{k}$, over the typical length scale $k_{0}^{-1}$ characterizing the wavelength. This enables us to ignore the LHS terms in the above equations, and to consider $\bar{\nabla}(\bar{k}\cdot\bar{x})\approx\bar{k}$ in them. After naming $\bar{p}=\bar{k}-\bar{\Gamma}$, the remaining system is just

\begin{eqnarray}
\bar{p}\times\bar{E}_{0}-\textbf{K} \,\bar{H}_{0}&=&0\label{eva10}\\
\bar{p}\times\bar{H}_{0}+\textbf{K} \,\bar{E}_{0}&=&0.\label{eva20}
\end{eqnarray}
These equations can by combined in order to obtain a single equation for (let us say) the electric field $\bar{E}_0$, which reads

\begin{equation}\label{eq:f1}
\textbf{K}^{-1}\left\lbrace \bar{p} \times \left[\textbf{K}^{-1} \left( \bar{p} \times \bar{E}_0\right) \right] \right\rbrace + \textbf{I}\bar{E}_0 = \bar{0}\,,
\end{equation}
where $\textbf{I}$ is the identity matrix. We have shown in Appendix \ref{ap1} how Eq. (\ref{eq:f1}) can be written as

\begin{equation}\label{eq:f3}
\left\lbrace\big(\bar{p} \otimes \bar{p}\big)\,\textbf{K} +\Big[ \det(\textbf{K}) -  \bar{p}^{\,\intercal} \textbf{K}\,\bar{p} \,
\Big] \textbf{I}\,\right\rbrace\bar{E}_0 = \bar{0}\,.
\end{equation}
In the same Appendix, it is shown how the vanishing of the determinant of the matrix defined by the quantity enclosed in braces in (\ref{eq:f3}) (a necessary condition required in order for non trivial solutions to exist), turns into the more diaphanous condition  

\begin{equation}\label{hamilton}
H\doteq\det(\textbf{K}) -\bar{p}^{\,\intercal} \textbf{K}\,\bar{p}=0.
\end{equation}
This is not only the dispersion relation (i.e., a constraint between the wave number $\bar{k}$ and the properties of the medium encoded in $\textbf{K}$ and $\bar{\Gamma}$), but also the starting point for ray tracing. As a matter of fact, $H$ can be viewed as a Hamiltonian governing the dynamics of light in the regime well described by  geometrical optics \cite{Sluijter1}-\cite{Mackay3}.

In view of the constraint $H=0$, the amplitude of the electric field results from the equation (\ref{eq:f3}) as

\begin{equation}\label{ampli2}
(\bar{p} \otimes \bar{p})\,\textbf{K}\,\bar{E}_0 = \bar{0}.
\end{equation}

\section{Exact results}\label{sec3}

\subsection{Hamilton equations for light rays in PT media}\label{sec3a}

Because $H(\bar{x}(t),\bar{k}(t))$ does not depends explicitly on $t$, it is conserved during the evolution (note that in (\ref{hamilton}), $\bar{p}$ is really a function of $\bar{k}$). The constancy of $H$ anywhere along a solution curve parametrized by $t$ means that

\begin{equation}\label{conecham}
\frac{dH}{dt}=0=\bar{\nabla}_{\bar{x}}H\cdot\frac{d\bar{x} }{dt}+\bar{\nabla}_{\bar{k}}H\cdot\frac{d\bar{k}}{dt},
\end{equation}
which implies Hamilton's canonical equations 
\begin{align}
\bar{\nabla}_{\bar{x}} H &=-\frac{d \bar{k}}{dt}\label{ham1gen1}\\
\bar{\nabla}_{\bar{k}}H &=\frac{d \bar{x}}{dt}\label{ham1gen2}.
\end{align}
Herein the shorthand $\bar{\nabla}_{\bar{v}}\equiv (\partial/\partial v_1,\partial/\partial v_2,\partial/\partial v_3) $ for $\bar{v}=(v_1,v_2,v_3)$ is adopted. However, $H$ is not only constant, but also identically null. This is actually a constraint that the evolution equations must preserve, even though it is not naturally contained in Eqs. (\ref{ham1gen1}) and (\ref{ham1gen2}); rather, the 6-dimensional curves $(\bar{x}(t),\bar{k}(t))$ coming from (\ref{ham1gen1}) and (\ref{ham1gen2}) must be further restricted to live in the hypersurface $H=0$ \footnote{The fact that (\ref{ham1gen1})-(\ref{ham1gen2}), plus $H=0$ is actually a constrained Hamiltonian system, as far as we know, seems to be overlooked in the literature. Many authors focus on solving (\ref{ham1gen1})-(\ref{ham1gen2}) (mainly numerically), without imposing the Hamiltonian constraint $H=0$, or at least, they do not explicitly mention at all doing otherwise}. In other words, Hamilton's equations are unable to determine the value of the constant ``energy level'' of the system, which is precisely what the dispersion relation (\ref{hamilton}) does.

Hamiltonian constrained systems emerges quite naturally in the context of field theories in which the dynamical variables have a gauge arbitrariness, as in classical electrodynamics or general relativity; as a matter of fact, constraints are indicators that the phase space is in some sense, too large, due to the freedom to perform gauge transformations. However, the \emph{Hamiltonian constraint} $H=0$ of Eq. (\ref{hamilton}) is of a very different nature. It comes to light because of the fact that the proposed fields (\ref{expEyH1}) and (\ref{expEyH2}) must obey Maxwell equations in the geometrical optics approximation.

We managed to find exact expressions for the system (\ref{ham1gen1}) and (\ref{ham1gen2}). Starting from (\ref{ham1gen1}) we have

\begin{equation}\label{apeb1}
\bar{\nabla}_{\bar{x}}H=\bar{\nabla}_{\bar{x}}(\det(\textbf{K}))-\bar{\nabla}_{\bar{x}}(\bar{p}^{\,\intercal}\textbf{K}\,\bar{p}).
\end{equation} 
The first term on the RHS can be evaluated directly using
\begin{equation}\label{apeb2}
\frac{\partial \det(\textbf{K})}{\partial x_{i}}=\det(\textbf{K}) \,tr(\textbf{K}\, \textbf{K}_{i}),
\end{equation}
where we have written $\textbf{K}_{i}=\partial\textbf{K}/\partial x_{i}$ (these are four matrices, whose components are obtained by differentiating the components of $\textbf{K}$ with respect to the coordinate $i$). Then
\begin{equation}\label{apeb3}
\bar{\nabla}_{\bar{x}}(\det(\textbf{K}))=\det(\textbf{K}) \,tr(\textbf{K} \,\textbf{K}_{i})\,\hat{e}_{i},
\end{equation} 
where $\hat{e}_{i}$ are the elements of the canonical basis on $\mathbb{R}^{3}$. The second term in (\ref{apeb1}) requires a bit more patience; using $\bar{p}=\bar{k}-\bar{\Gamma}$, it follows that 
\begin{equation}\label{apeb4}
\bar{p}^{\,\intercal}\textbf{K}\,\bar{p}=\bar{k}^{\,\intercal}\textbf{K}\,\bar{k}-2\bar{k}^{\,\intercal}\textbf{K}\,\bar{\Gamma}+\bar{\Gamma}^{\,\intercal}\textbf{K}\,\bar{\Gamma}.
\end{equation}
The different contributions read
\begin{eqnarray}
\bar{\nabla}_{\bar{x}}(\bar{k}^{\,\intercal}\textbf{K}\,\bar{k})&=&\frac{\partial(\bar{k}^{\,\intercal}\textbf{K}\,\bar{k})}{\partial x_{i}}\,\hat{e}_{i}=\bar{k}^{\,\intercal}\textbf{K}_{i}\,\bar{k}\,\hat{e}_{i}, \nonumber\\
\bar{\nabla}_{\bar{x}}(\bar{k}^{\,\intercal}\textbf{K}\,\bar{\Gamma})&=&\frac{\partial(\bar{k}^{\,\intercal}\textbf{K}\,\bar{\Gamma})}{\partial x_{i}}\,\hat{e}_{i}=\bar{k}^{\,\intercal}(\textbf{K}_{i}\,\bar{\Gamma}+\textbf{K}\,\bar{\Gamma}_{i})\,\hat{e}_{i},\nonumber\\
\bar{\nabla}_{\bar{x}}(\bar{\Gamma}^{\,\intercal}\textbf{K}\,\bar{\Gamma})&=&\frac{\partial(\bar{\Gamma}^{\,\intercal}\textbf{K}\,\bar{\Gamma})}{\partial x_{i}}\,\hat{e}_{i}=\bar{\Gamma}^{\,\intercal}(2\textbf{K}\,\bar{\Gamma}_{i}+\textbf{K}_{i}\,\bar{\Gamma})\,\hat{e}_{i},\nonumber
\end{eqnarray}
where $\bar{\Gamma}_{i}=\partial\bar{\Gamma}/\partial x_{i}$ and we extensively have used the fact that $\partial\bar{k}/\partial x_{i}=0$. Adding the corresponding terms and rewriting the result in terms of $\bar{p}$, we have

\begin{equation}\label{apeb8}
\bar{\nabla}_{\bar{x}}(\bar{p}^{\,\intercal}\textbf{K}\,\bar{p})=\bar{p}^{\,\intercal}(\textbf{K}_{i}\,\bar{p}+2\textbf{K}\,\bar{p}_{i})\,\hat{e}_{i},
\end{equation}
where $\bar{p}_{i}=\partial\bar{p}/\partial x_{i}$. Finally, using (\ref{apeb3}), (\ref{apeb8}) and (\ref{apeb1}), Eq. (\ref{ham1gen1}) results

\begin{equation}
\frac{d \bar{k}}{dt}=\left[ \bar{p}^{\,\intercal}(\textbf{K}_{i}\,\bar{p}+2\textbf{K}\,\bar{p}_{i})-\det(\textbf{K})\,tr(\textbf{K}^{-1}\textbf{K}_{i})\right]\hat{e}_{i}\,, \label{hamiltonmom}
\end{equation}
It is important to bear in mind that the expression within brackets in (\ref{hamiltonmom}) is an $i$-dependent scalar, and summation in $i$ is understood.

On the other hand, in regard to the remaining Hamilton equation (\ref{ham1gen2}), things are much easier; from the definition of $H$ in (\ref{hamilton}), it is quite obvious that (\ref{ham1gen2}) results

\begin{equation}
\frac{d \bar{x}}{dt}=2 \textbf{K}\bar{p},\label{hamiltoncoor}
\end{equation}

Having derived (\ref{hamiltonmom}) and (\ref{hamiltoncoor}), we can easily incorporate the constraint $H=0$ into the system; we only need to put $\det(\textbf{K}) =\bar{p}^{\,\intercal} \textbf{K}\,\bar{p}$ as coming from (\ref{hamilton}), into (\ref{hamiltonmom}). Hence, the final dynamical equations are 

\begin{eqnarray}
\frac{d \bar{x}}{dt}&=&2 \textbf{K}\bar{p},\label{hamiltoncoorfin}\\
\frac{d \bar{k}}{dt}&=&\bar{p}^{\,\intercal}\Big[ [\textbf{K}_{i}-tr(\textbf{K}^{-1}\textbf{K}_{i})\,\textbf{K}]\,\bar{p}+2\textbf{K}\,\bar{p}_{i}\Big]\hat{e}_{i}\,.\label{hamiltonmomfin}
\end{eqnarray}
These are the exact dynamical equations that will serve as a starting point for geometrical-optics, ray-tracing analysis in PT media.

\subsection{Optical analogue of the G\"odel spacetime}\label{sec3b}

The matrix $\textbf{K}$ introduced in (\ref{relgama}), obtained for the specific metric considered in this work (\ref{metgSI}), is:

\begin{equation}\label{eq:K}
\textbf{K}=\frac{\sqrt{2}}{2}\,\mathrm{diag}\left(e^{\sqrt{2}  \omega x_1 },2\,e^{-\sqrt{2}  \omega x_1 }, e^{\sqrt{2}  \omega x_1 }\right)\,,
\end{equation}
and its determinant is just:
\begin{equation}\label{detgama}
\det(\textbf{K})=\frac{\sqrt{2}}{2}\, e^{\sqrt{2}  \omega x_1} \,.
\end{equation}
The vector $\bar{\Gamma}$ in (\ref{elotrogama}) results:
\begin{equation}
\bar{\Gamma} =
\begin{pmatrix}
0 \\
e^{\sqrt{2} \omega  x_1}\\
0 
\end{pmatrix}
\,.
\end{equation}
Hence, we have
\begin{equation}\label{hamiltonpa}
\bar{p}^{\,\intercal} \textbf{K}\,\bar{p}=
\frac{e^{\sqrt{2}  \omega x_1 }}{\sqrt{2}}\Big[p_{1}^2 + 2\, p_{2}^2 \,e^{-2 \sqrt{2} \omega x_1} + p_{3}^2\Big],
\end{equation}
Using (\ref{detgama})-(\ref{hamiltonpa}), and the fact that $\bar{p}=\bar{k}-\bar{\Gamma}$, we can write the Hamiltonian (\ref{hamilton}) in terms of $\bar{x}$ and $\bar{k}$ as: 
\begin{equation}\label{hamiltonfin}
H=\frac{e^{\sqrt{2} \omega x_1 }}{\sqrt{2}}\Big[1 - k_{1}^2 - 2 \,e^{-2\sqrt{2} x_1 \omega} \big(k_{2} - e^{\sqrt{2} \omega x_1}\big)^2 - k_{3}^2\Big].
\end{equation}
A proper change of coordinates defined by:
\begin{align}\label{vchangeadim}
(t,x_1,x_2,x_3) \mapsto (\sqrt{2}\omega)^{-1}(\tau,u_{1},u_{2}, u_{3})
\end{align}
transforms the space-time coordinates $(t, \bar{x})$ into the non-dimensional  $(\tau, \bar{u})$. The dispersion relation in the new variables obtained by setting $H=0$ is:
\begin{equation}\label{reldis}
k_{1}^2 + 2 \,e^{-2 u_{1}} \big(k_{2} - e^{u_{1}}\big)^2 + k_{3}^2 = 1\,.
\end{equation}
Due to the fact that $u_2$ and $u_3$ are cyclic coordinates, $k_2$ and $k_3$ are constants of motion. The relation (\ref{reldis}) provides the link between coordinates and momenta in such a way that the fields (\ref{mostro}) are solutions of Maxwell's equations in the geometrical optics realm. 

However, in order for the Hamiltonian formalism to be fruitful, we need to assure the absence of evanescent modes, which are fields with a complex-valued wave vector $\bar{k}$. In addition to having $k_{2},k_{3}\in \mathbb{R}$, the real character of the vector $\bar{k}$ is assured from (\ref{reldis}) if:
\begin{equation}\label{reldis4}
u_1^{-}\leq u_{1}\leq u_1^{+}, \,\,\,\,u_1 ^{\pm}=\ln\big[(2\pm\sqrt{2})k_{2}\big],\,\,\,\,k_{2}>0.
\end{equation}
The limit case in (\ref{reldis4}) leads to a wave vector of the form $\bar{k}=(0,k_{2},0)$. 
This means that propagating (non evanescent) modes exist only in a region of the medium given by (\ref{reldis4}), for all $u_{2}$ and $u_{3}$. Nonetheless, this restricted $u_{1}$-space depends on the (strictly) positive constant of motion $k_{2}$, which can be arbitrarily large. Few paragraphs below we shall find further restrictions on the components of the wave vector $\bar{k}$.

Let us now proceed constructively from the unconstrained Hamilton's equations (\ref{hamiltonmom}) and (\ref{hamiltoncoor}). For the case under consideration, after the application of the chain rule, they adopt the form:
\begin{align}
\frac{d\bar{k}}{d\tau}&=-\sqrt{2}\left(\resizebox{0.6cm}{!}{$\frac{e^{u_{1}}}{2}$} \big( 1 + k_{1}^2 + k_{3}^2\big) -\,k_{2}^2\,e^{-u_{1}},0,0\right)
\label{ham1}\\[0.5cm]
\frac{d\bar{u}}{d\tau}&=\sqrt{2}\Big( e^{u_{1}}\,k_{1},2\big(k_{2}\,e^{-u_{1}}-1\big), e^{u_{1}}\,k_{3} \Big)\label{ham2}.
\end{align}
These first order differential equations are coupled and can be explicitly solved for each vector component.  From now on, we shall write $\dot{(...)}=d(...)/d\tau$. Let us begin solving them for the $1^{st}$ vector component.
From (\ref{ham2}) we get
\begin{equation}\label{ecfif3}
k_{1}=\frac{e^{-u_{1}}}{\sqrt{2}}\dot{u}_{1},
\end{equation}
Performing the derivative of $k_{1}$ in (\ref{ecfif3}) with respect to the variable $\tau$:
\begin{equation}
\dot{k}_{1}=\frac{e^{ -u_1}}{\sqrt{2}}(\ddot{u}_{1}-\dot{u}_{1}^2).\label{ecfif3prim2}
\end{equation}
With the help of (\ref{ecfif3}) and (\ref{ecfif3prim2}) we can write down the first component of (\ref{ham1}) in the form of a second order, nonlinear differential equation for $u_{1}(\tau)$:
\begin{equation}\label{ecfif4}
\ddot{u}_{1}-\resizebox{0.25cm}{!}{$\frac{1}{2}$}\dot{u}_{1}^2+(1+k_{3}^2)\, e^{2u_{1}}-2k_{2}^2=0.
\end{equation}
As mentioned in the previous section, Hamilton's equations do not guarantee the Hamiltonian constraint (dispersion relation), so neither does (\ref{ecfif4}). In order to include the information coming from $H=0$, let us use (\ref{ecfif3}) in the dispersion relation (\ref{reldis}) with the purpose of solving for $\dot{u}_{1}$. This leads to

\begin{equation}\label{ecfif5}
- \frac{\dot{u}_{1}^2}{2}=e^{2u_{1}}(1+k_{3}^2-4k_{2}\,e^{-u_1}),
\end{equation}
which can be replaced in (\ref{ecfif4}) to finally obtain 

\begin{equation}\label{ecfif6}
\ddot{u}_{1}+2e^{2u_{1}}(1+k_{3}^2-2k_{2}e^{-u_1})-2k_{2}^2=0.
\end{equation}
It is straightforward to show that this equation can be obtained directly from the constrained Hamiltonian system (\ref{hamiltoncoorfin})-(\ref{hamiltonmomfin}).

Fortunately, equation (\ref{ecfif6}) can be solved exactly. Nevertheless, its resolution will bring over further restrictions on some components of $\bar{k}$, otherwise the solution would be a complex-valued function of $\tau$; it can be shown that, provided
\begin{equation}\label{eq:condkykz}
\frac{1}{2} < k_2, \,\,\,\,0 \leq \vert k_3 \vert < 1 - \frac{1}{(2 k_2)^2}\,,
\end{equation}
the real solution of (\ref{ecfif6}) is

\begin{equation}
u_1(\tau)= \ln \Bigg[k_2 \Big( 1 + \resizebox{1.3cm}{!}{$\sqrt{\frac{1 - k_3^2}{2}}$} \cos \big[ 2 k_2 (\tau + C_{1})\big]\Big)^{\!\!-1} \!\Bigg], \label{eq:Solxtau}
\end{equation}
where $ C_{1}$ is an integration constant. Note that the admissible values of $k_{2}$ according to (\ref{eq:condkykz}) are slightly more restrictive that the ones coming from the requirement of absence of evanescent modes, Eq. (\ref{reldis4}). This means that the smaller set (\ref{eq:condkykz}) is enough, not only for having proper propagating modes, but also for assuring real-valued trajectories. Once the constants of motion $k_{2}$ and $k_{3}$ are selected to fulfill (\ref{eq:condkykz}), the (non-constant) $k_{1}$ is obtained from the dispersion relation (\ref{reldis}) by means of (\ref{eq:Solxtau}). Besides, it can be easily verified that when $k_3$ is fixed at its limiting value $k_{3}=0$, and the cosine function is valued at its upper or lower bound, then (\ref{eq:Solxtau}) reaches the limiting values $u_1^+$ and $u_1^-$ displayed at (\ref{reldis4}). This means that the function (\ref{eq:Solxtau}) verifies all the constraints involved for any value of $k_{2}$ and $k_{3}$ compatible with (\ref{eq:condkykz}).   

The remaining components of the trajectory are obtained at once from $u_{1}$. The second and third components of (\ref{ham2}) imply
\begin{eqnarray}
u_2(\tau)&=&2\sqrt{2}\int \Big(k_{2}\,e^{-u_1 (\tau)}-1\Big)d\tau\label{ecfif41},\\
u_3(\tau)&=&\sqrt{2}\,k_{3}\int e^{u_1(\tau)}d\tau \label{ecfif42},
\end{eqnarray}
which in turn lead us to

\begin{align}
u_2(\tau)&= \resizebox{1.2cm}{!}{$\frac{\sqrt{1 - k_3^2}}{k_2}$} \, \sin \big[2 k_2 (\tau + C_{1})\big] + C_2, \label{eq:Solytau}\\[0.3cm] 
u_3(\tau)&= \resizebox{1.3cm}{!}{$\frac{ 2\,k_{3} }{\sqrt{1+k_3^2}}$}\arctan \left(\tilde{k}_{3} \tan \left[ k_2 (\tau + C_{1})\right] \right) +\! C_{3},\label{eq:Solztau}
\end{align}
where $C_{2}$, $C_{3}$ are integration constants, and $\tilde{k}_{3}=\left(\sqrt{2}-\sqrt{1-k_3^2}\right)/\sqrt{1+k_3^2}$. On the other hand, while $k_2$ and $k_3$ are constants, $k_1$ evolves as 
\begin{equation}\label{k1tau}
k_1(\tau) = \sqrt{1-k_3^2} \sin \big[ 2 k_2 (\tau + C_{1})\big].
\end{equation}

It must be noted that $C_1$ has no physical meaning and can be zeroed without any loss of generality.

The purpose of the next section is to carefully study these results.

\section{Structure of the light paths}\label{sec3c}

\subsection{Closed light orbits}

A remarkable topological feature shared by all trajectories in this peculiar material is that they all form simple closed curves when projected onto the $(u_1,u_2)$-plane. This follows from the fact that, under the homeomorphism  $(u_1,u_2) \longmapsto (U_1,U_2)$, with 
    \begin{equation*}
        U_1 = \sqrt{\frac{2}{1-k_3^2}}\left(k_2 e^{-u_1}-1\right), \,\,\,\,\,\,
        U_2 = \frac{k_2}{\sqrt{1-k_3^2}} u_2, 
    \end{equation*}
all trajectories form unit circles in the $(U_1,U_2)$-plane (this comes from (\ref{reldis}), (\ref{eq:Solytau}) and (\ref{k1tau})). Also, these orbits have a period, measured along the parameter $\tau$, of $\pi / k_2$. 

As a reference, we will trace light rays together with G\"odel's CNC, which is the boundary of the region in which causal pathologies occur. Therefore we must know how CNCs and CTCs (defined in G\"odel's spacetime by $r=r_{0}$ and $r>r_{0}$, respectively) look like in the $(u_1,u_2)$-coordinates (\ref{vchangeadim}). From (\ref{change1}) we get
	\begin{equation}\label{cond2}
	\sin^2\phi=1-\Big[\frac{\exp (u_1)-\cosh(2r)}{\sinh(2r)}\Big]^2,
	\end{equation}
	and using (\ref{change2}) we have instead
	\begin{equation}\label{cond3}
	\sin^2\phi=\frac{u_2^2 \exp (2u_1)}{2\sinh^2(2r)}.
	\end{equation}
	Combining (\ref{cond2}) and (\ref{cond3}) we obtain the desired relation between $u_1$ and $u_2$,
	
	\begin{equation}\label{condctcs}
	u_2^2 = -2 - 2 e^{-2u_1} + 4 e^{-u_1}\cosh(2r).
	\end{equation}
	In particular, G\"odel's CNC defined by $r=r_{0}$ verifies 
	\begin{equation}\label{cond4}
	u_2^2 = -2 - 2 e^{-2u_1} + 12e^{-u_1}.
	\end{equation}

\begin{figure}
    \centering
    \includegraphics[scale=0.45]{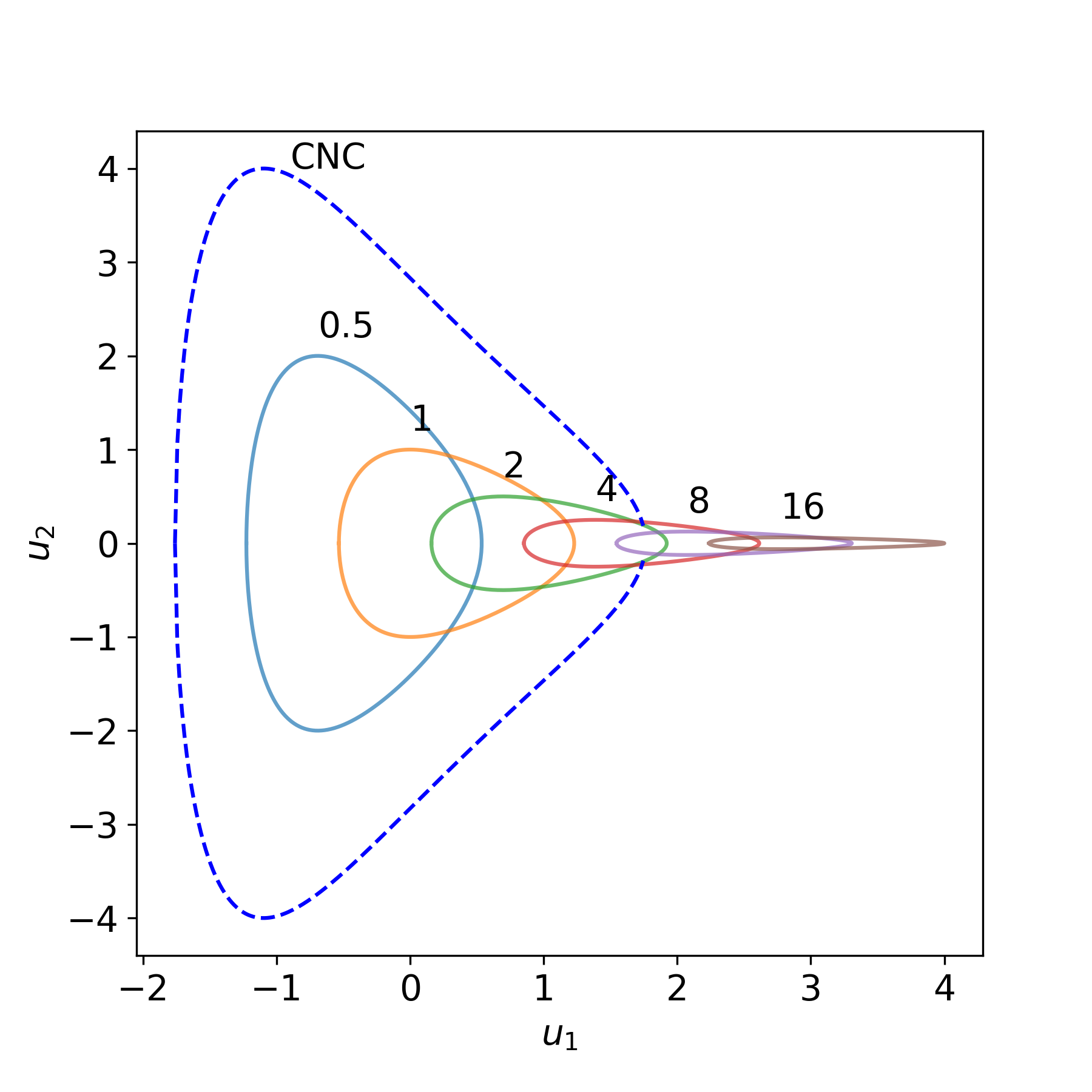}
    \caption{Closed orbits confined to the $(u_1,u_2)$-plane, for different values of $k_{2}$. The curve associated to Gödel's CNC is also shown for reference (dashed line).}
    \label{fig:cl_orb}
\end{figure}

We focus now on the particular case $k_{3}=0$, where the trajectories (\ref{eq:Solxtau})-(\ref{eq:Solztau}) are properly closed (i.e., they are themselves their own projection on the $(u_1,u_2)$-plane). In Fig.~\ref{fig:cl_orb}, five closed orbits are depicted. Starting from the limit curve corresponding to $k_2=0.5$ (note that, according to Eq.~\eqref{eq:condkykz}, $k_2=0.5$ does not belong to the parameter space), and letting $k_2$ to grow exponentially, we see that the location along the $u_1$-axis moves linearly, as expected from the logarithmic nature of the limits for $u_1$ (cf. Eq.~\eqref{reldis4}). Meanwhile, the centered interval covered by the cycles in the $u_2$-axis scales inversely proportional to $k_2$. From (\ref{eq:Solxtau}), it is clear that the span over the $u_1$-axis is determined by the values of both $k_2$ and $k_3$. 
Given an arbitrary anchor point in the $(u_2,u_3)$-plane, trajectories unroll towards a boundary in the medium space as $k_2$ goes to infinity. On the other hand, there exists a boundary for the minimum value $u_1$ can attain, given by $-\ln [2 + \sqrt{2}]$; this value comes from taking $k_2=1/2$ in Eq.  (\ref{eq:Solxtau}).

As mentioned, the orbit corresponding to the CNC in Gödel's universe (see Eq. (\ref{cond4})), was depicted as a reference in a dashed line. Note that, contrary to what happens in (3+1)-dimensional spacetime, this curve is not a limit curve of any sort; closed light rays in the material medium are not limited to exist in the region defined by the interior of the Gödel's CNC, as it is the case for null geodesics in Gödel's spacetime \cite{Tedescos},\cite{Tedescos2}. In turn, closed non-evanescent light rays in the medium are allowed to circle not only in the interior of the Gödel's CNC, but in the exterior as well. This includes trajectories intersecting Gödel's CNC at two points, without having any causal pathology. This will be further clarified below.

\subsection{Spiral Trajectories}

When $k_3=0$, there is no net displacement of the electromagnetic fields, as evidenced by the closed orbit solutions of Fig.~\ref{fig:cl_orb}. Whenever $k_3$ is not equal to zero, orbits unfold along the $u_3$-axis (cf. Eq.~\eqref{eq:Solztau}).

Now, we turn our attention to the analysis of spatial dispersion in this context. To formulate such analysis we begin by considering quasi-plane waves with different wave vectors but fixing a common initial position in the $(u_1,u_2,u_3)$ coordinates. This can be thought of as launching a wave packet from a given point in space. 

A characteristic feature of our solutions (\ref{eq:Solxtau})-(\ref{eq:Solztau}) is that, while there is freedom to choose an arbitrary initial position for $u_2$ and $u_3$, the initial position of $u_1$ is constrained by the wave numbers $k_2$ and $k_3$, or vice versa. From Eq.~\eqref{eq:Solxtau} we notice that, if we conventionally fix $u_1(0)=u_1^0$, then $k_2$ and $k_3$ are connected by the relation 

\begin{equation}
k_{3}=\sqrt{1-2(k_{2}e^{-u_1^0}-1)^2}.
\label{eq:k2_const}
\end{equation}

Also, from Eqs.~\eqref{eq:k2_const} and \eqref{eq:condkykz}, we find that $0 \leq |k_3| < k_3^{\mathrm{sup}}$, where this supremum for the magnitude that $k_3$ can attain arises as a solution of
\begin{equation}\label{eq:k3sup}
  4 e^{2 u_1^0} (1- k_3) \left(1+2\sqrt{\frac{1-k_3^2}{2}}+\frac{1-k_3^2}{2} \right)=1.  
\end{equation}

\begin{figure}
    \centering
    \includegraphics[scale=0.45]{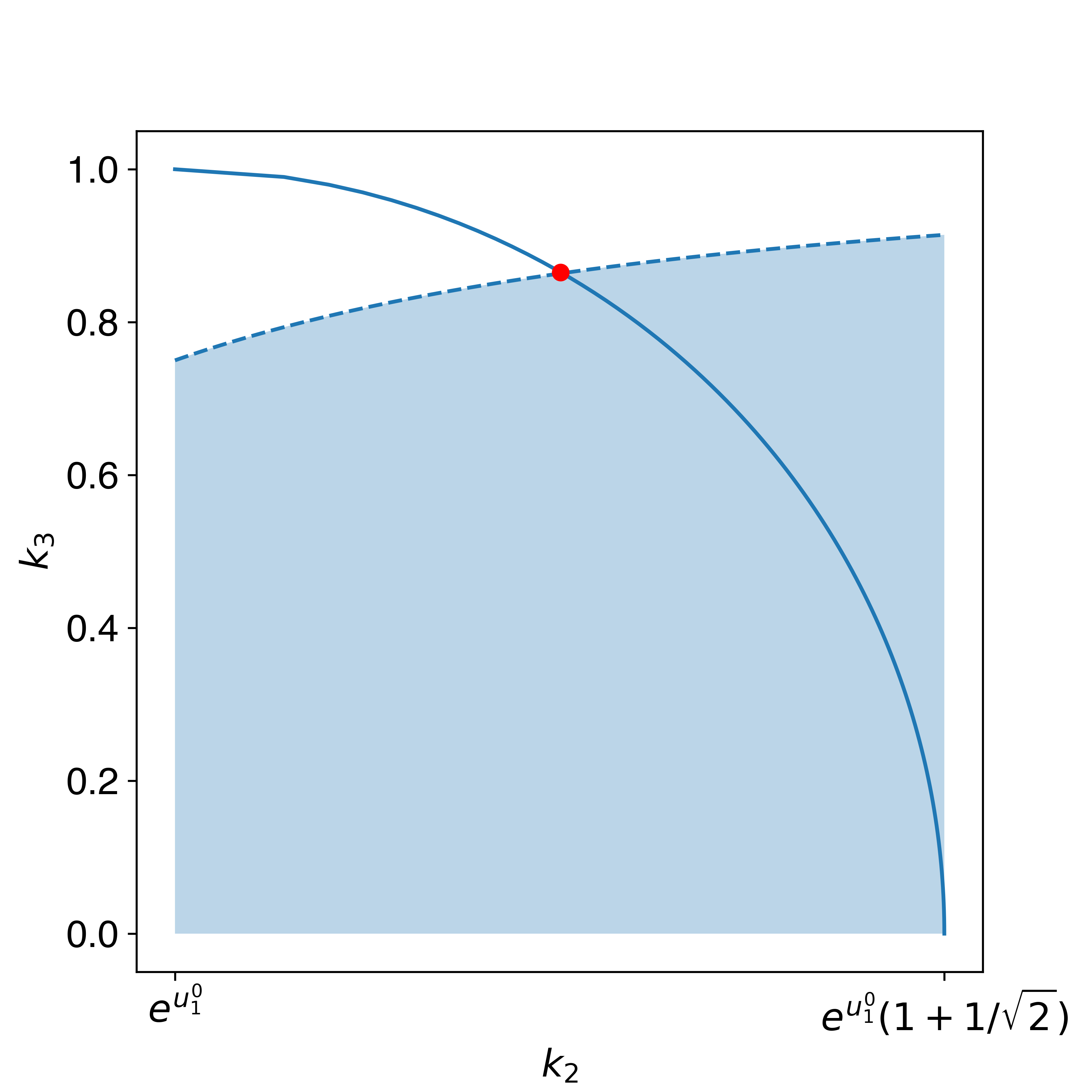}
    \caption{Equation~\eqref{eq:k2_const} (solid blue line) together with the constraint of Eq.~\eqref{eq:condkykz} (blue shaded area). The red dot corresponds to $(k_2^{\mathrm{inf}},k_3^{\mathrm{sup}})$. Despite the limits for $k_2$ are expressed in general, the current figure corresponds to $u_1^0=0$.}
    \label{fig:k3sup}
\end{figure}

Even though Eq.~\eqref{eq:k3sup} has no closed form solution, in Fig.~\ref{fig:k3sup} the limits for $k_3$ can be easily envisaged. Besides, since $k_2^{\mathrm{inf}}$, the $k_2$ value corresponding to $k_3^{\mathrm{sup}}$, verifies  $e^{u_1^0} \leq k_2^{\mathrm{inf}} \leq e^{u_1^0}(1+1/\sqrt{2})$, we can conclude that in order to attain an unlimited net displacement condition, it is necessary to make $u_1^0$ grow without bound. 

\begin{figure}
    \centering
    \includegraphics[scale=0.5]{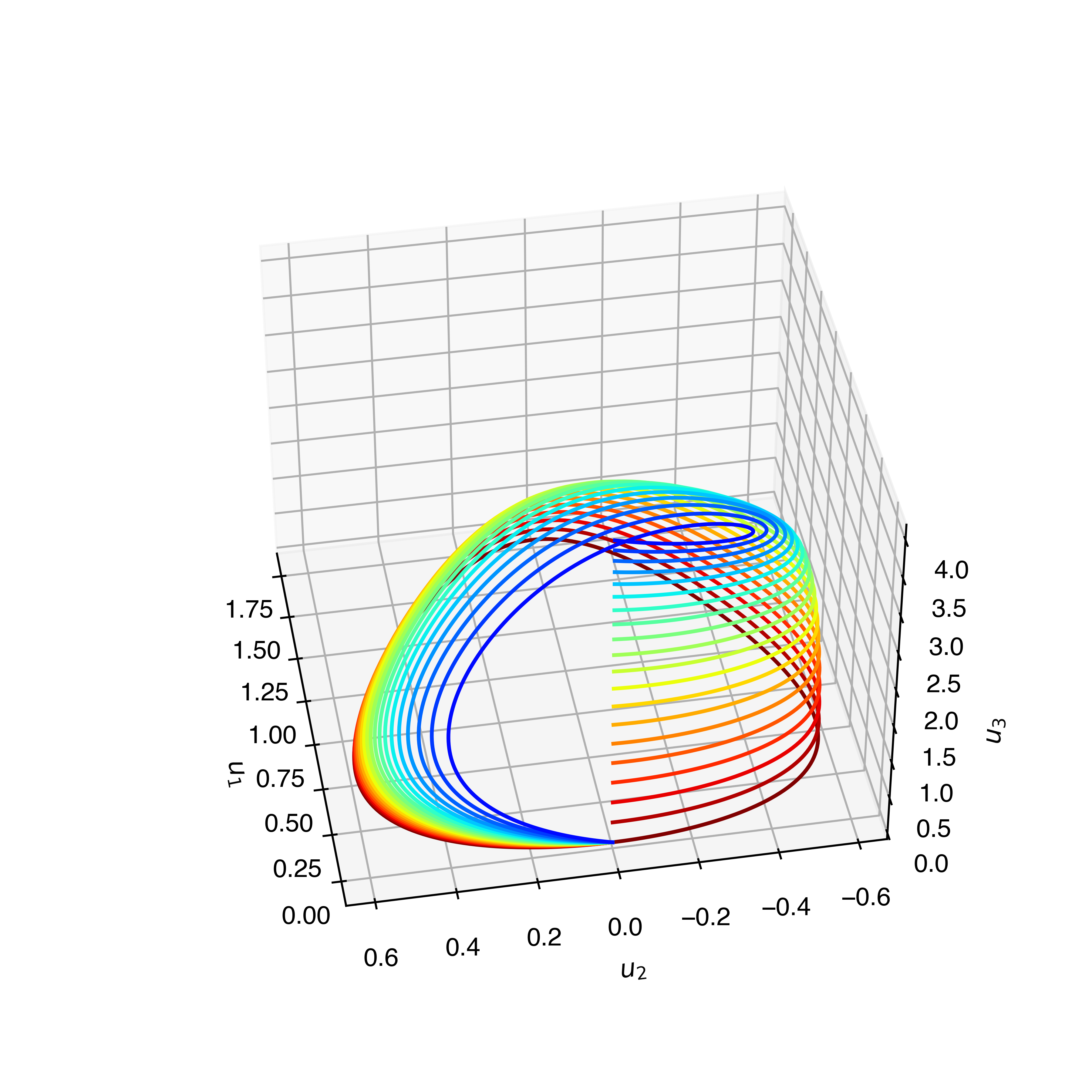}
    \caption{Trajectories for the different \emph{components} of a wave packet launched at $(0,0,0)$. The third component of the wavevector, $k_3$, ranges linearly from 0 (dark red) to $k_3^{\mathrm{sup}}$ (dark blue).}
    \label{fig:chromatic}
\end{figure}

Let us proceed to draw trajectories departing from a given point in space, say $\bar{u}(0)=(u_1^0,u_2^0,u_3^0)$. Firstly, we note that, as a consequence of Eq.~\eqref{eq:k2_const}, information can neither travel in closed orbits nor following a single trajectory. As a matter of fact, any information-bearing field, and therefore a wave packet having non-zero bandwidth, will consist of a superposition of monochromatic waves, each having a different wave vector $\bar{k}$; hence, by the condition imposed by Eq.~\eqref{eq:k2_const}, each monochromatic quasi-plane wave component will evolve from the initial point $\bar{u}_0$ following a different trajectory, as seen in Fig.~\ref{fig:chromatic}. 

\begin{figure}
    \centering
    \includegraphics[scale=0.5]{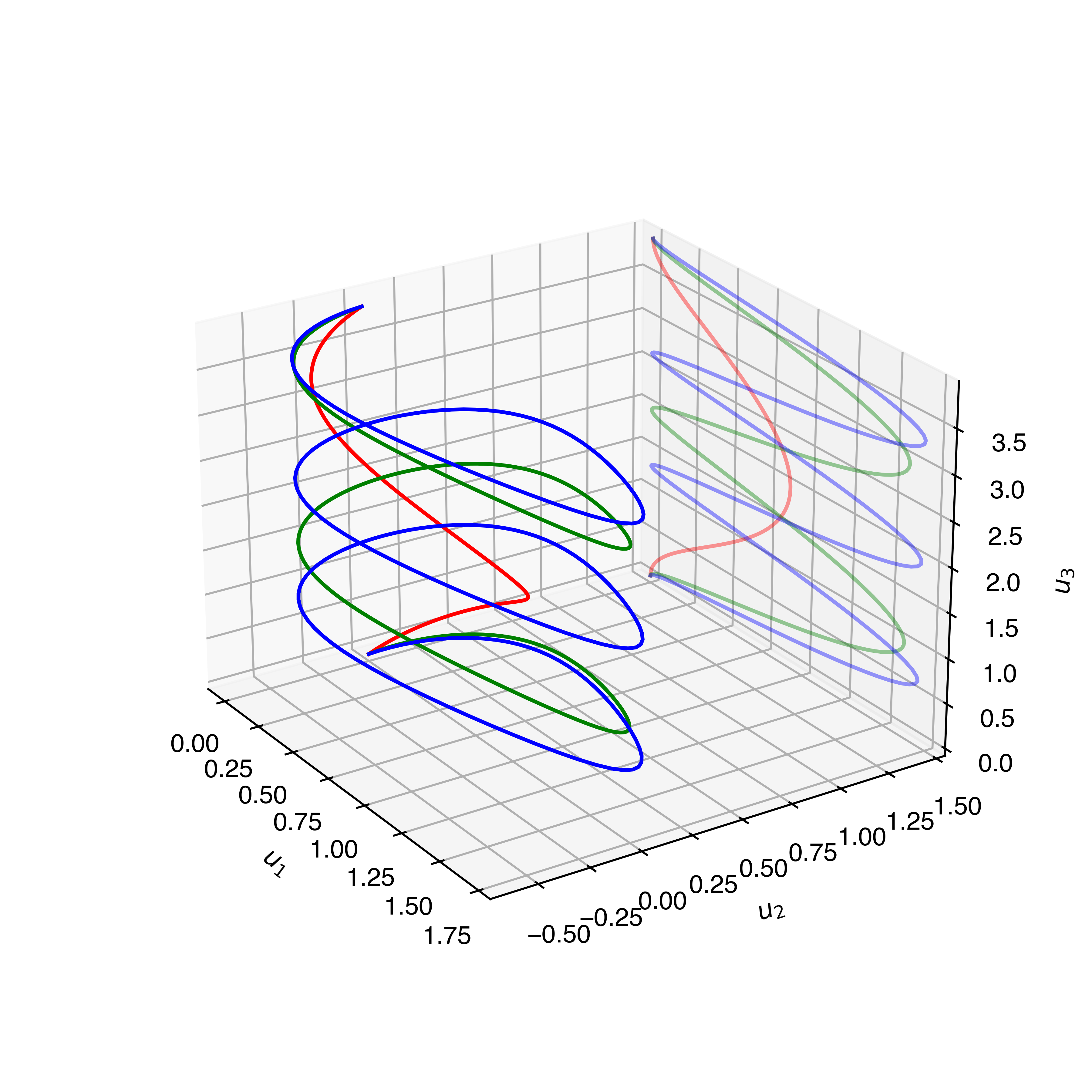}
    \caption{Periodical refocusing of three monochromatic quasi-plane waves that comply with Eq.~\eqref{eq:CondRefocus}, for $m/\ell=2$ (solid green) and $m/\ell=3$ (solid blue), with respect to $k_3=0.8$ (solid red). A projection onto the $(u_1,u_3)$-plane is shown for better visualization.}
    \label{fig:refocusing}
\end{figure}

Nevertheless, a \emph{refocusing} phenomenon can be achieved at certain prescribed points in space, and for a countable subset of harmonics of any given wave packet. To further analyze this property let us assume, without any loss of generality, that a wave packet departs from $\bar{u}^0=(0,0,0)$. As it was already mentioned, every trajectory has a closed, and therefore periodic, projection onto the $(u_1,u_2)$-plane, the period being $\pi/k_2$. Besides, trajectories unwind in the $u_3$-axis direction obeying Eq.~\eqref{eq:Solztau}. If $u_3(\tau)$ were linear in $k_2 \tau$ it would be straightforward to find the relation between the different harmonics such that they periodically meet (or refocus), also periodically, in the $u_3$-axis. However, even though the function $\arctan \left(\tilde{k}_{3} \tan (k_2 \tau) \right)$ appearing in (\ref{eq:Solztau}) wiggles periodically about the linear function $k_2 \tau$ for $\tilde{k}_{3} \neq 1$ (for $\tilde{k}_{3} = 1$ it is exactly $k_2 \tau$), we realize that for constant $A$, $\tilde{k}_{3}$ and $k_{2}$ we have

\begin{equation}
A \,\arctan\left(\tilde{k}_{3}\tan(k_{2} \tau)\right) = A\, k_{2}\, \tau,\,\,\,\, \mbox{if}\,\,\,\,\tau = n\, \frac{\pi}{2 k_{2}},\,\,\,\,  n \in \mathbb{Z}\,.\nonumber
\end{equation}
Then the relation that any two wave vectors $\bar{k}^{(1)},\bar{k}^{(2)}$ must meet in order to produce the refocusing, is given by the following equation linking their corresponding second and third components
\begin{equation}\label{eq:CondRefocus}
\frac{k_3^{(1)}}{k_3^{(2)}}\frac{\sqrt{1+\Big(k_3^{(2)}\Big)^2}}{\sqrt{1+\Big(k_3^{(1)}\Big)^2}} = \frac{m}{\ell}, \,\,\, \,\,m, \ell \in \mathbb{Z}, 
\end{equation}
together with the condition given by Eq.~\eqref{eq:k2_const}. In this way, the corresponding quasi-plane wave components for each wavevector will refocus periodically at 

\begin{equation}
u_3^{(1)}(2 \pi \ell n/k_2^{(1)})=u_3^{(2)}(2 \pi m\, n/k_2^{(2)}),\,\,\, \,\,n \in \mathbb{Z},  \nonumber
\end{equation}
where $u_3^{(i)}(\tau)$ refers to the solution for a given wavevector $\bar{k}^{(i)}$. See Fig.~\ref{fig:refocusing} for an example of three such refocusing quasi-plane waves.

Finally, notice that there is a limit in the vertical net displacement given by the maximum value that $k_3$ can adopt, also posing a limit in how small the projections of the trajectories in the $(u_1,u_2)$-plane can be. 

\subsection{Power flux}\label{secpowerflux}

We are particularly interested in analyzing the power flux along ray trajectories. Given that the quasi-plane wave approximation applies, in a sufficiently small neighbourhood of any point in space such that the non-uniform medium can be approximated as uniform, the time-averaged Poynting vector can be obtained, as usual, by doing \cite{Mackay4}

\begin{equation} \label{eq:tavgpoynting}
    \left< \bar{S} (\tilde{\omega}, \bar{k}) \right> _t = \frac{1}{2} \mathrm{Re} \{ \bar{E}(\tilde{\omega}, \bar{k}) \times \bar{H}^*(\tilde{\omega}, \bar{k})  \}
\end{equation}
where $\bar{E}(\tilde{\omega}, \bar{k})$ and $\bar{H}(\tilde{\omega}, \bar{k})$ are the complex-valued phasors corresponding to the Fourier representation of $\bar{E}(t)$ and $\bar{H}(t)$, respectively, at a given point in space. For a quasi-plane wave such as those of Eqs.~\eqref{expEyH1}-\eqref{expEyH2}, $\bar{E}(\tilde{\omega}, \bar{k})$ and $ \bar{H}(\tilde{\omega}, \bar{k})$ coincide with $\bar{E}_0$, $\bar{H}_0$.

A solution for the electromagnetic field amplitude is not an outcome of the formalism at use but, nevertheless, we can infer the value of \eqref{eq:tavgpoynting} at any point by setting an arbitrary initial/boundary condition $\bar{E}_0$ complying with the condition of Eq.~\eqref{ampli2}, which is equivalent to choosing $\bar{E}_0$ such that $\bar{p}^{\; \intercal}\mathbf{K} \bar{E}_0=0$, i.e., $\bar{E}_0$ must be orthogonal to $\mathbf{K} \bar{p}$ (see (\ref{ape8})). This condition, for the Gödel analogue translates into
\begin{equation}\label{ecaplitud}
k_{1}E_{01}+2e^{-2u_1}(k_{2}-e^{u_1})E_{02}+k_{3}E_{03}=0.
\end{equation}

By using Eqs.~\eqref{eva10}, (\ref{ape4}) and vector cross product identities, we may express Eq.~\eqref{eq:tavgpoynting} conveniently in terms of the electric field $\bar{E}_0$ as
\begin{equation} \label{eq:tavgPoyn2}
    \left< \bar{S} \right> _t = \frac{\bar{E}_0^{\; \intercal} \mathbf{K} \bar{E}_0^*}{2 \mbox{det}(\mathbf{K})}  \mathbf{K} \bar{p},
\end{equation}
where we also used the fact that $\bar{p}^{\; \intercal}\mathbf{K} \bar{E}_0=0$, and $\mathbf{K}^*=\mathbf{K}$. 

We are only interested in the magnitude $\|\left< \bar{S} \right> _t\|$ since, by virtue of Eq.~\eqref{hamiltoncoorfin} we can see that the direction of $\left<\bar{S}\right>_t$ is always parallel to the tangent of the trajectories in the $(u_1,u_2,u_3)$ space. Nonetheless, replacing $\mathbf{K}$ and $\bar{p}$ for the Gödel analogue, we have that Eq.~\eqref{eq:tavgpoynting} yields
\begin{equation} \label{eq:tavgPoynGodel}
    \left< \bar{S} \right> _t =  f(E_{0i},u_{1}) \, 
      \left[ 
    \begin{array}{c}
         k_1 \\
         2 (k_2 - e^{u_1}) e^{-2 u_1}\\
         k_3
    \end{array}
    \right],
\end{equation}
where
\begin{equation}
f(E_{0i},u_{1})=\frac{e^{u_1}}{2\sqrt{2}}(|E_{01}|^2 + 2|E_{02}|^2 e^{-2u_1} + |E_{03}|^2).
\end{equation}

Before conducting an analysis of the magnitude of the Poynting vector over light paths, we should choose the electric field at each point over the trajectories. In addition to the condition that $\bar{E}_0$ must be orthogonal to $\mathbf{K} \bar{p}$, we add the extra constraint
\begin{equation}
    \label{eq:constEuni}
    |E_{01}|^2 + |E_{02}|^2 + |E_{03}|^2 =1.
\end{equation}

This choice obeys to the following phenomenology: the medium under consideration is linear in the electromagnetic fields, then it does not have different regimes for different field magnitudes. Hence it is useful to fix $\|\bar{E}_0\|$. 

\begin{figure}
    \centering
    \includegraphics[width=0.45\textwidth]{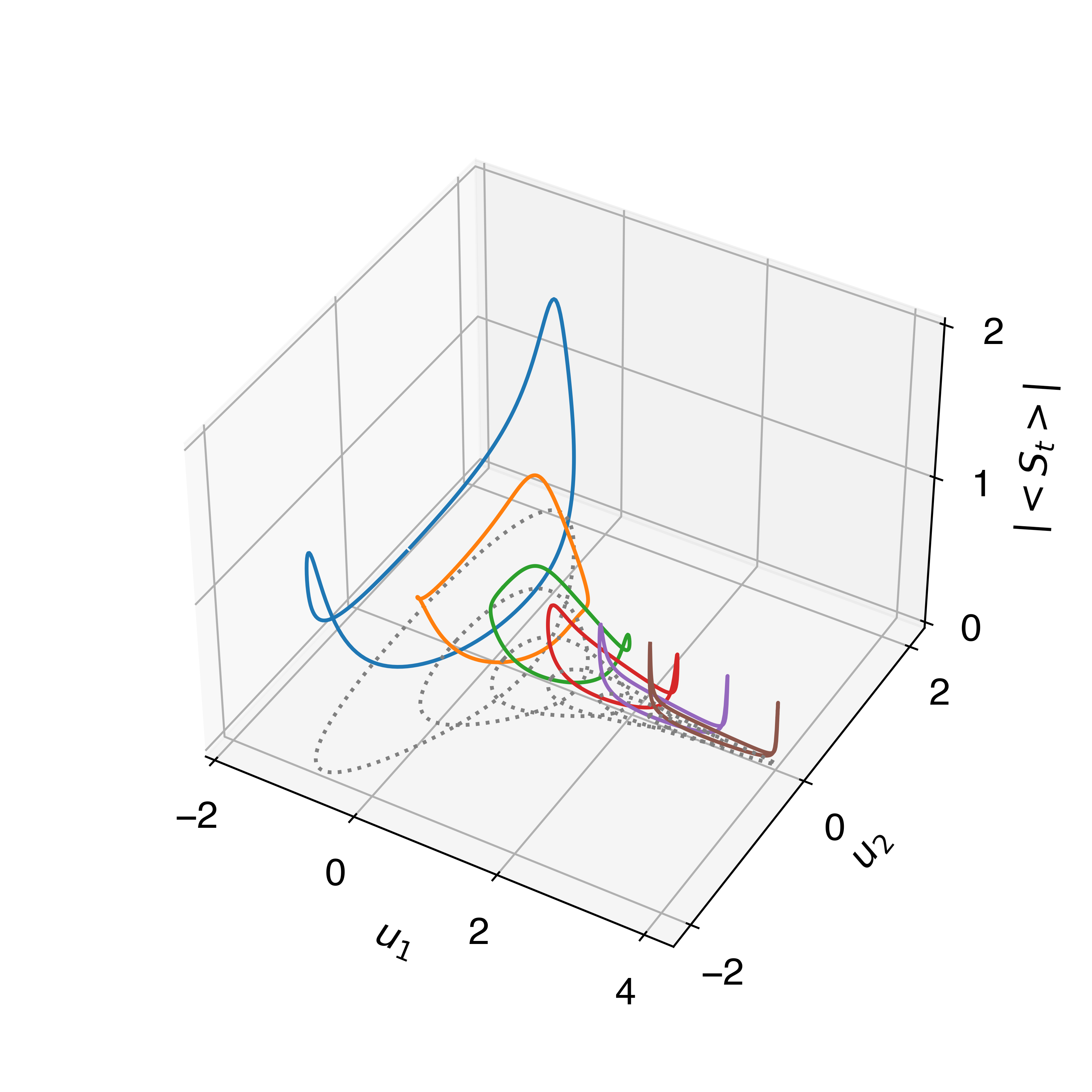}
    \caption{Power flux (vertical axis) when $k_3=0$ for electric fields parallel to the $(u_1,u_2)$ plane for the same values of $k_2$ as in Fig.~\ref{fig:cl_orb}. The orbits are shown for reference in dotted gray lines.}
    \label{fig:Poyn1}
\end{figure}

We start by analysing the simplest case, that of the closed trajectories, i. e., $k_3=0$. To further simplify the analysis, we also consider electric fields confined to the $(u_1,u_2)$ plane, i.e., $E_{03}=0$. Then, from Eqs.~\eqref{ecaplitud} and  \eqref{eq:constEuni}
\begin{equation} \label{eq:Eorth1}
\bar{E}_0 = \pm \sqrt{\frac{\alpha^2}{1+\alpha^2}} (1, \alpha^{-1}, 
0)
\end{equation}
where $\alpha :=k_1^{-1}[2 e^{-2 u_1}(e^{u_1}-k_2)]$. In this case, we obtain the power flux magnitudes shown in Fig.~\ref{fig:Poyn1}, where both plus and minus sign in Eq.~\eqref{eq:Eorth1}, i. e., an electric field pointing inwards or outwards the orbits, give the same result. 

\begin{figure}
    \centering
    \includegraphics[width=0.45\textwidth]{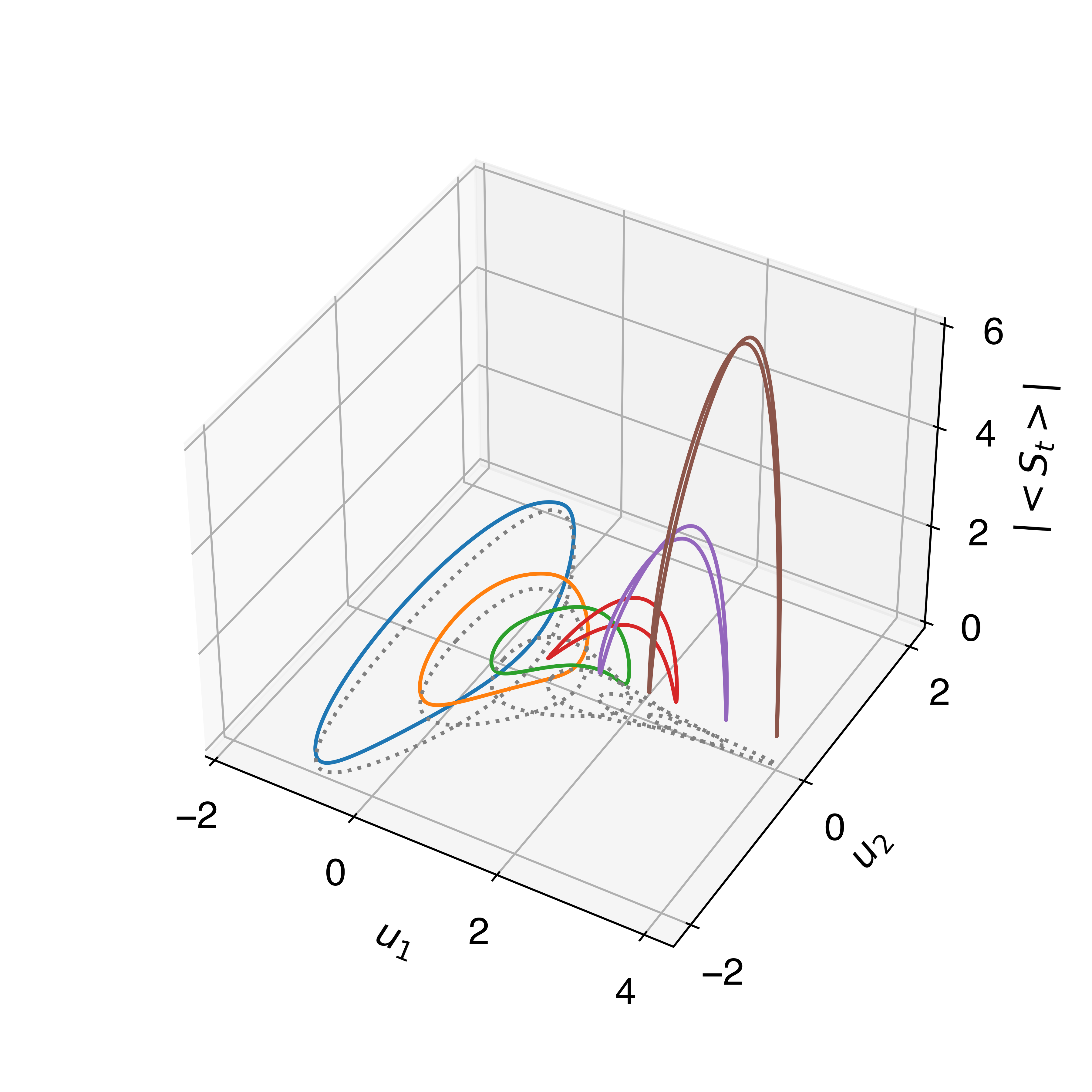}
    \caption{Power flux (vertical axis) when $k_3=0$ for electric fields parallel to the $u_3$ axis for the same values of $k_2$ as in Fig.~\ref{fig:cl_orb}. The orbits are shown for reference in dotted gray lines.}
    \label{fig:Poyn1_2}
\end{figure}

Also, for the closed orbits, it is easy to analyze the effect of an electric field in an orthogonal direction to that of the plane where the orbits evolve, by simply considering $\bar{E}_0=(0,0,1)$. Results for this scenario are shown in Fig.~\ref{fig:Poyn1_2}.

\begin{figure}
    \centering
    \includegraphics[width=0.45\textwidth]{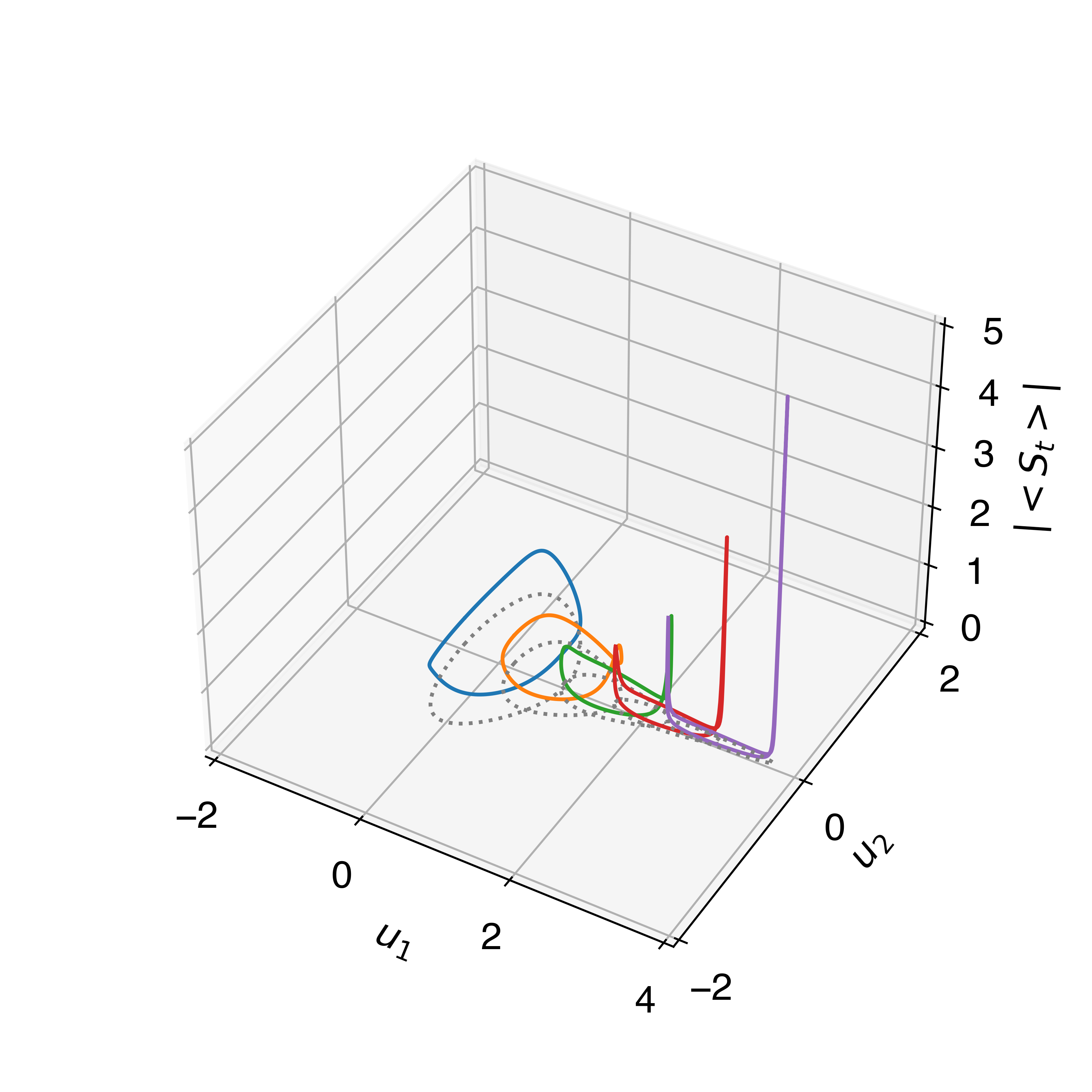}
    \caption{Power flux (vertical axis) when $k_3=0.5$ for electric fields parallel to the $(u_1,u_2)$ plane for $k_2=1,2,4,8,16$. The projection of the orbits are shown for reference in dotted gray lines.}
    \label{fig:poyn2}
\end{figure}

In Fig.~\ref{fig:poyn2}, we explore the effect of having a net $u_{3}$-displacement on the power flux magnitude. Hence, we set $k_3 \neq 0$ for the same electric field constraint as in the first case. Therefore, we set $k_3=0.5$ and, to comply with Eq.~\ref{eq:condkykz}, we exclude $k_2 = 0.5$ from the considered set.

\begin{figure}
    \centering
    \includegraphics[width=0.45\textwidth]{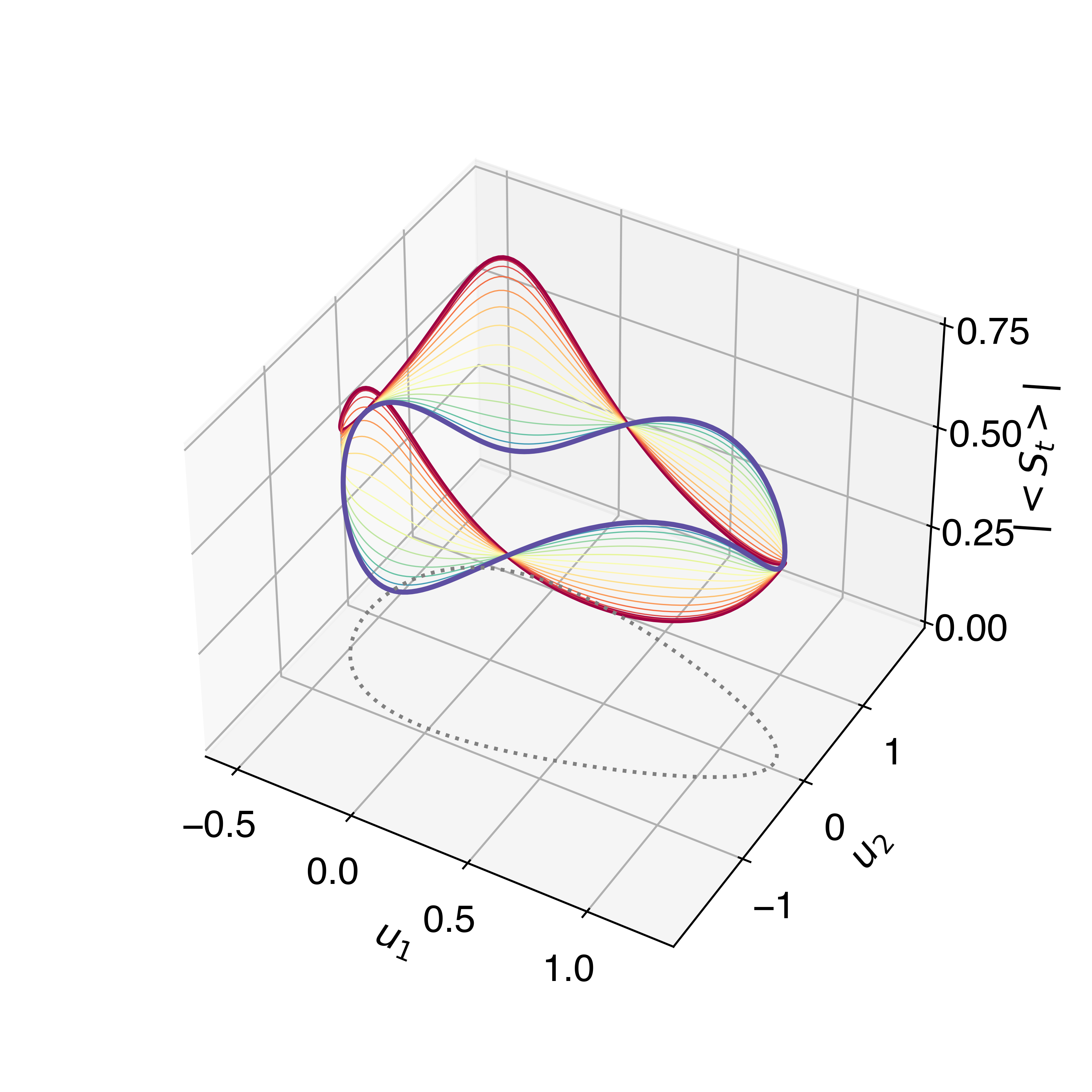}
    \caption{Power flux for different initial electric fields, rotated as indicated in Eq. (\ref{erot}), with $k_2=1$ and $k_3=0$. The angle $\theta$ sweeps from 0 radians (purple) to $\pi/2$ radians (dark blue).}  
    \label{fig:poyn_rotatingE0k3zero}
\end{figure}

\begin{figure}
    \centering
    \includegraphics[width=0.45\textwidth]{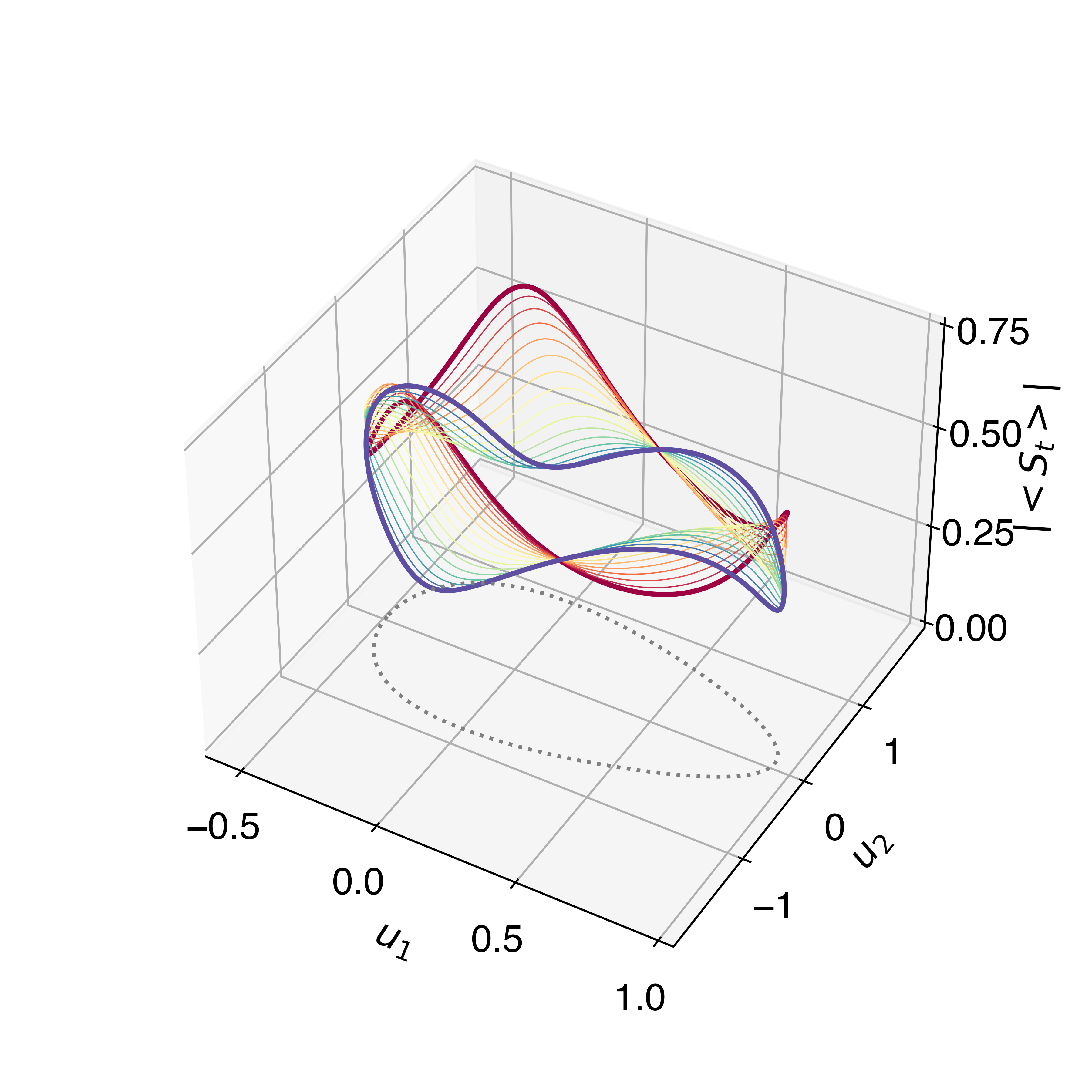}
    \caption{Power flux for different initial electric fields, rotated as indicated in Eq. (\ref{erot}), with $k_2=1$ and $k_3=0.5$. The angle $\theta$ sweeps from 0 radians (purple) to $\pi/2$ radians (dark blue).}
    \label{fig:poyn_rotatingE0k3nonzero}
\end{figure}

We might also investigate the behavior of the power flux magnitude when a unit length $\bar{E}_0$ vector rotates around $\mathbf{K} \bar{p}$. We exhibit how this magnitudes vary for the case $k_2=1$, $k_3=0$ in Fig.~\ref{fig:poyn_rotatingE0k3zero} and $k_3=0.5$ in Fig.~\ref{fig:poyn_rotatingE0k3nonzero}. It can be noticed that the distribution of the power flux has two clearly distinctive modes for a $\pi/2$ rotation of the electric field. Rotations about an axis, namely $\bar{r} = \mathbf{K} \bar{p}/\| \mathbf{K} \bar{p} \|$, is performed considering an initial $\bar{E}_0$ orthogonal to $\bar{r}$ and then evaluating 

\begin{equation} \label{erot}
\bar{E}_0^\theta = \cos(\theta) \bar{E}_0 + \sin(\theta) (\bar{E}_0 \times \bar{r}),
\end{equation} 
where $\bar{E}_0$ is always taken according to Eq.~\eqref{eq:Eorth1}. In Figs. (\ref{fig:poyn_rotatingE0k3zero}) and (\ref{fig:poyn_rotatingE0k3nonzero}), we depict the power flux coming from the different electric fields parameterized by $\theta$ as in Eq. (\ref{erot}).

\section{Closing Remarks}
\label{sec:concl}
Geometrical optics in Plebanski-Tamm media is governed by the Hamiltonian system characterized by Eqs. (\ref{hamiltoncoorfin}) and (\ref{hamiltonmomfin}), which were written down in exact form for the first time in this work. These equations automatically incorporate the Hamiltonian constraint $H=0$ (see Eq. (\ref{hamilton})), which is no other than the dispersion relation linking the properties of these unusual media with the wavevector $\bar{k}$. As an interesting working example, we exactly solved Hamilton equations for light rays in the case in which the Plebanski-Tamm medium corresponds to the optical analogue of G\"odel's universe. It is worth of mention that G\"odel's metric can be written in the static coordinates of Eq. (\ref{metgSI}), which render the analysis possible. This is ultimately due to the fact that G\"odel's spacetime does not describe an expanding universe, but a stationary, rotating one instead. 

Among the strange phenomena occurring in the G\"odel analogue PT medium, we should highlight the chromatic behavior of plane waves, which we proceed to summarize:

a) Light is able to circle round and round endlessly, provided $k_{3}=0$. The closed trajectories exist in both the interior and the exterior regions defined by G\"odel's closed null curve (see Fig. (\ref{fig:cl_orb})), and some trajectories actually cross it showing no causal anomalies of any sort.

b) If $k_{3}\neq0$ the trajectories spiral round and round incessantly along the entire range of the $u_{3}$ coordinate. If a wave packet is launched from a certain initial position, each quasi-plane monochromatic component evolves differently according to the magnitude of its wavevector (see Fig. (\ref{fig:chromatic})).

c) If $k_{3}\neq0$ some trajectories periodically \emph{refocus}. This is a sort of \emph{multi-imaging} produced by the fact that there are an infinite number of trajectories joining two arbitrary points separated by a non null value of the $u_{3}$ coordinate, because of the spiral structure of the light paths (see Fig. (\ref{fig:refocusing})). This is of course reminiscent of the behavior of null geodesics in the (3+1) G\"odel spacetime. An alleged inhabitant of such a peculiar material will experience the world through a show of multiple images and colors coming out of every existing object.

It is clear that these results are possible because of the fact that the G\"odel-like PT medium is geodesically complete in the Riemannian sense; the trajectories are allowed to exist for all values of the parameter $\tau$, and they are not interrupted because of geometrical obstructions, as curvature singularities.

Besides points (a)-(c) above, we managed to study also the power flux arising by imposing different initial conditions for the electric field amplitudes, obtaining the results condensed in Figs. (\ref{fig:Poyn1})-(\ref{fig:poyn_rotatingE0k3nonzero}) of section \ref{secpowerflux}.

We conclude our study by envisaging future developments along the following lines of research:

1) The study of spacetime singularities by knowing the behavior of light rays in the 3-metric constituting the anisotropic medium. In particular, to establish relations between the singularity theorems in (3+1)-dimensional spacetime and the completeness of light ray trajectories in the optical medium. 

2) The analysis concerning the formation and stability of Cauchy horizons in (3+1)-dimensional spacetime by monitoring the behavior of light in the analogue medium, that is to say, the impact that the causal violations in spacetime have on the structure of light rays propagating in the medium. Concretely, to study the analogue of spacetimes having closed null geodesics (see, e.g., \cite{Boston}). 

3) The quantization of the constrained Hamiltonian system (\ref{hamiltoncoorfin})-(\ref{hamiltonmomfin}) and a thorough characterization of the constraints therein involved, as well as a proper identification of the genuine physical degrees of freedom.

\bigskip

Finally, we should briefly comment on the experimental feasibility of the concepts and ideas evoked in this work. In the particular case of interest, as seen in the governing constitutive relations (\ref{eq:ecconstitD}) and (\ref{eq:ecconstitB}), the magnetic field induces electrical polarization and vice versa, creating so a bi-anisotropic (impedance-matched) material verifying $\epsilon_{ij}(\bar{x})=\mu_{ij}(\bar{x})$, via the action of the matrix $\textbf{K}(\bar{x})$ defined in Eq.~(\ref{relgama}). Impedance-matched media are certainly evasive to actual practical implementations, but some progress towards the realization of metamaterial prototypes was made nonetheless. For instance, microwave devices were built for mimicking the 2D version of Maxwell´s fish eye \cite{Meta5}, \cite{Meta6}, a \emph{perfect imaging} system envisaged by Maxwell itself during the very early days of electromagnetism \cite{Meta4}. It would be interesting to inquire on what kind of microwave device could reproduce the behavior of light studied in this work.

%Se puede decir algo sobre la “Chiralidad” \\

%Segun leí por ahí (por ahi= wikipedia :\ ) si se cumple

%$$ \xi =-\zeta ^{T}=-i\kappa ^{T}} $$
%adonde en nuestro caso $\xi$ seria el rotor y $\zeta$ seria menos el %rotor,  y $\kappa$ el "chiral tensor".

%y sobre (en mi opinión lo más controversial) el hecho de que $\epsilon$ y $\mu$ son iguales... 

\subsection*{Acknowledgments}
FF is a member of Carrera del Investigador Científico (CONICET), and his work is supported by CONICET and Instituto Balseiro.

\appendix
\section{On the equation for the electric field amplitude}\label{ap1}

Let us work out a bit Eq. (\ref{eq:f1}). First, let us remember that $\det(\textbf{K})\, \textbf{K}^{-1}= (\mathrm{cof}(\textbf{K}))^{\intercal}$, where $\mathrm{cof}(\textbf{K})$ is the cofactor matrix of $\textbf{K}$. But $\textbf{K}$ is a symmetric matrix, then transposing the last equation we easily get

\begin{equation}\label{ape1}
\det(\textbf{K}) \,\textbf{K}^{-1}=\mathrm{cof}(\textbf{K}).
\end{equation}
Another vector identity will be extremely useful; for $n\times1$ vectors $\bar{a}$ and $\bar{b}$ and a $n\times n$ matrix $\textbf{A}$ we have

\begin{equation}\label{ape2}
(\textbf{A}\bar{a})\times(\textbf{A}\bar{b})=\mathrm{cof}(\textbf{A})(\bar{a}\times\bar{b}).
\end{equation}
If we make the choice
\begin{equation}\label{ape3}
\textbf{A}=\textbf{K}, \,\,\,\,\bar{a}=\bar{p},\,\,\,\,\bar{b}=\textbf{K}^{-1}(\bar{p}\times\bar{E}_{0}),     
\end{equation}
the first term of the LHS of (\ref{eq:f1}) can be written as

\begin{equation}\label{ape4}
\textbf{K}^{-1}\left\lbrace \bar{p} \times \left[\textbf{K}^{-1} \left( \bar{p} \times \bar{E}_0\right) \right] \right\rbrace =\frac{\textbf{K}\,\bar{p} \times (\bar{p}\times \bar{E}_0)}{\det(\textbf{K})}.
\end{equation}
We can eliminate from (\ref{ape4}) the double vector product by using $\bar{a}\times \bar{b}\times \bar{c}=\bar{b}\,(\bar{a}\cdot\bar{c})-\bar{c}\,(\bar{a}\cdot\bar{b})$. This lead us to 
\begin{equation}\label{ape5}
\frac{\textbf{K}\,\bar{p} \times (\bar{p}\times \bar{E}_0)}{\det(\textbf{K})}=\frac{\bar{p}\left[\textbf{K}\,\bar{p}\cdot\bar{E}_0\right]-\bar{E}_0\left[\textbf{K}\,\bar{p}\cdot\bar{p}\right]}{\det(\textbf{K})}.
\end{equation}
The quantities inside the brackets in (\ref{ape5}) are scalars; we can transpose them and obtain

\begin{eqnarray}
\textbf{K}\,\bar{p}\cdot\bar{p}&=&\left(\textbf{K}\,\bar{p}\cdot\bar{p}\right)^{\intercal}=\bar{p}^{\intercal}\textbf{K}\,\bar{p}\label{ape6} \\ 
\textbf{K}\,\bar{p}\cdot\bar{E}_0&=&\left(\textbf{K}\,\bar{p}\cdot\bar{E}_0\right)^{\intercal}=\bar{p}^{\intercal}\textbf{K}\,\bar{E}_0=\bar{p}\cdot\textbf{K}\,\bar{E}_0\,,\label{ape7}
\end{eqnarray}
where we have used $\textbf{K}=\textbf{K}^{\intercal}$. Finally, with the help of (\ref{ape7}) we can write the first term of the RHS of (\ref{ape5}) as

\begin{equation}\label{ape8}
\bar{p}\left[\textbf{K}\,\bar{p}\cdot\bar{E}_0\right]=(\bar{p}\otimes\bar{p})\textbf{K}\,\bar{E}_0\,.
\end{equation}
Here, the Kronecker (tensor) product of two vectors $\bar{c}$ and $\bar{a}$ is defined as usual according to
\begin{equation}\notag
\bar{c}\otimes\bar{a}=\left(
\begin{array}{ccc}
c_1\\
c_2\\
c_3
\end{array}
\right)
\otimes
\left(
\begin{array}{ccc}
a_1 & a_2 & a_3 
\end{array}
\right)
=
\left(
\begin{array}{ccc}
c_1 a_1 & c_1 a_2 & c_1 a_3 \\
c_2 a_1 & c_2 a_2 & c_2 a_3 \\
c_3 a_1 & c_3 a_2 & c_3 a_3 \\
\end{array}
\right).
\end{equation}

Gathering all the results, adding the second term $\textbf{I}\bar{E}_0$ of (\ref{eq:f1}), and factorizing $\bar{E}_0$ to the right, we finally obtain the desired Eq. (\ref{eq:f3}).

Let us proceed now to proof (\ref{hamilton}). Assume for the moment that $H=\det(\textbf{K}) -\bar{p}^{\,\intercal} \textbf{K}\,\bar{p}\neq0$. Dividing (\ref{eq:f3}) by $H$ we get $\textbf{Z}\,\bar{E}_{0}$, where

\begin{equation}\label{calcint}
\textbf{Z}=\emph{I}+H^{-1}(\bar{p}\otimes\bar{p}) \textbf{K}=\textbf{I}+\bar{p}\otimes (H^{-1}\textbf{K}\,\bar{p}).
\end{equation}
In order to find non trivial solutions we need $\det(\textbf{Z})=0$. We make use of Sylvester's theorem in the form
\begin{equation}\label{silvestre}
\det(\textbf{Z})=1+H^{-1}(\textbf{K}\,\bar{p})^{\,\intercal}\bar{p}=1+H^{-1}\bar{p}^{\,\intercal}\textbf{K}\,\bar{p}\,.
\end{equation}
However, the condition $\det(\textbf{Z})=0$, i.e.,

\begin{equation}\label{silvestre2}
1+H^{-1}\bar{p}^{\,\intercal}\textbf{K}\,\bar{p}=1+\frac{\bar{p}^{\,\intercal}\textbf{K}\,\bar{p}}{\det(\textbf{K}) -\bar{p}^{\,\intercal} \textbf{K}\,\bar{p}}=0,
\end{equation}
only can be fulfilled if $\det(\textbf{K})=0$, which is not true in view of the very definition of $\textbf{K}$ (see. Eq. (\ref{relgama})). Hence, $H=0$ and (\ref{hamilton}) holds.

\end{document}